# A high-throughput analysis of ovarian cycle disruption by mixtures of aromatase inhibitors

Frederic Y. Bois[1*], Nazanin Golbamaki-Bakhtyari[1], Simona Kovarich[2], Cleo Tebby[1], Henry A. Gabb[3], Emmanuel Lemazurier[1].

[1] INERIS, DRC/VIVA, Parc ALATA, BP 2, 60550 Verneuil en Halatte, France.

[2] S-IN Soluzioni Informatiche Srl, Via Ferrari 14, I-36100 Vicenza – Italy.

[3] School of Information Sciences, University of Illinois at Urbana-Champaign, Champaign, Illinois, USA

* Corresponding author; Frederic Y. Bois, INERIS, DRC/VIVA, Parc ALATA, BP 2, 60550 Verneuil en Halatte, France; phone: 33-344-234-385; email: frederic.bois@ineris.fr

# KEYWORDS

Endocrine disruption, high-throughput risk assessment, mixtures, exposure, female reproductive cycle, toxicokinetics, women's health

# ACKNOWLEDGEMENTS

The research leading to these results received funding from the European Union's 7th Framework Program and Cosmetics Europe through grant agreement 266835 (COSMOS), the European Union's Horizon 2020 research and innovation program under grant agreement 633172 (Euromix) and the French Ministry for the Environment (PRG190). We thank the US EPA Computational Toxicology Research team for its help in using ToxCast and ExpoCast. The authors declare they have no financial interests linked to this research. They thank the reviewers for their helpful comments.

# ABBREVIATIONS

| | |
|---|---|
| CNS: | Central nervous system |
| CV: | Coefficient of variation |
| CYP: | Cytochrome P450 |
| E2: | Estradiol |
| EDC: | Endocrine disrupting chemical |
| FSH: | Follicle stimulating hormone |
| GnRH: | Gonadotropin-releasing hormone |
| LH: | Luteinizing hormone |
| P4: | Progesterone |
| PK: | Pharmacokinetics |
| QSAR: | Quantitative structure-activity relationship |

# ABSTRACT

***Background***: Combining computational toxicology with ExpoCast exposure estimates and ToxCast assay data gives us access to predictions of human health risks stemming from exposures to chemical mixtures.

***Objectives***: To explore, through mathematical modeling and simulations, the size of potential effects of random mixtures of aromatase inhibitors on the dynamics of women's menstrual cycles.

***Methods***: We simulated random exposures to millions of potential mixtures of 86 aromatase inhibitors. A pharmacokinetic model of intake and disposition of the chemicals predicted their internal concentration as a function of time (up to two years). A ToxCast aromatase assay provided concentration-inhibition relationships for each chemical. The resulting total aromatase inhibition was input to a mathematical model of the hormonal hypothalamus-pituitary-ovarian control of ovulation in women.

***Results***: Above 10% inhibition of estradiol synthesis by aromatase inhibitors, noticeable (eventually reversible) effects on ovulation were predicted. Exposures to individual chemicals never led to such effects. In our best estimate, about 10% of the combined exposures simulated had mild to catastrophic impacts on ovulation. A lower bound on that figure, obtained using an optimistic exposure scenario, was 0.3%.

***Conclusions***: These results demonstrate the possibility to predict large-scale mixture effects for endocrine disrupters with a predictive toxicology approach, suitable for high-throughput ranking and risk assessment. The size of the effects predicted is consistent with an increased risk of infertility in women from everyday exposures to our chemical environment.

# INTRODUCTION

Concern is growing worldwide over the negative human health and environmental impacts of chemical pollutants that can interfere with the production, metabolism, and action of natural hormones, the so-called endocrine-disrupting chemicals (EDCs). In humans, EDCs have been linked to reproductive disorders (Sweeney et al. 2015), abnormal or delayed development in children (Schug et al. 2015), and changes in immune function (Rogers et al. 2013) or cancer (Birnbaum and Fenton 2002). Exposure to mixtures of EDCs may result in effects that can depart from mere summation (Kortenkamp 2007), and human subgroups (*e.g.*, women) may not be sufficiently protected against mixtures of EDCs by current regulatory limits (Kortenkamp 2014).

Each menstrual cycle in women involves hormonal regulation of follicular growth and maturation resulting in ovulation of a single oocyte (Falcone and Hurd 2013). The cycle is controlled by coordinated stimulations and inhibitions along the hypothalamus-pituitary-ovarian axis. GnRH, secreted by the hypothalamus, stimulates the secretion of gonadotropins (FSH and LH) by the anterior pituitary gland. Those in turn regulate the secretion of ovarian hormones, such as estradiol (E2) or progesterone (P4). Exposures to EDCs interfering directly or indirectly with any of these hormones can eventually induce infertility or other pathological outcomes. Aromatase is critical because it irreversibly converts testosterone to E2 and androstenedione to estrone, maintaining the dynamic balance between androgens and estrogens.

The objective of the current work is to explore predictively the effects of exposure to large scale (*i.e.*, potentially real-life) mixtures of aromatase inhibitors on the dynamics of menstrual cycling in women. We input exposure estimates from ExpoCast (Wambaugh et al. 2013) and biological effect data from ToxCast (Dix et al. 2007) to coupled pharmacokinetic (PK) and ovarian cycle models. This provided a quantitative mechanistic link between exposure to mixtures of EDCs and their potential adverse effects on the menstrual cycle in women. We compared the expected effects of exposures to single EDCs, as usually considered by risk assessment provisions in different regulations, to estimated effects of cumulative and concurrent exposures.

## METHODS

### *Workflow overview*

The overall computational workflow is pictured on Figure S1 in Supplemental Material. Briefly, after selecting the chemicals of interest, we sampled millions of random mixtures of chemicals using the exposure estimates provided by ExpoCast (Wambaugh et al. 2013). Both constant and time-varying exposure scenarios of an adult woman were considered. A pharmacokinetic model of intake and disposition was then used to estimate the blood concentration (over two years) for each chemical present in each mixture. The resulting aromatase activity inhibition was estimated using the Hill's dose-response model parameters provided by ToxCast (Dix et al. 2007). A mathematical model of the hypothalamus-pituitary-ovarian hormonal events (based on Chen and Ward 2013) was used to predict the levels of E2, P4, and other quantities characterizing the ovarian cycle, for a reference cycle and following exposure to the mixtures generated. Monte Carlo sampling (Bois et al. 2010) was used to propagate uncertainties in exposure, kinetics, and dose-response relationships up to ovarian cycle perturbation.

### *Databases, chemical selection, and mixture sampling*

ExpoCast (Wambaugh et al. 2013) provides exposure estimates, with measures of uncertainty, for 1936 chemicals. Those exposure estimates were obtained using far-field, mass-balance human exposure models (USEtox and RAIDAR).

In ToxCast (11 December 2013 release), the *Tox21-aromatase-inhibition* assay is a cell-based assay measuring CYP19A1 (aromatase) gene activity *via* a fluorescent protein reporter gene. Chemicals acting on aromatase *mRNA* synthesis, degradation or translation, or on aromatase itself should give positive results in this assay (Chen et al. 2015). For each chemical *x* assayed, ToxCast provides the geometric mean and standard error for the parameter values $AC_{50,x}$, $W_x$, $B_x$ and $T_x$ of a Hill function fitted to the concentration-inhibition data (scaled using the positive and negative controls' data):

$$\%Inhibition \;=\; B_x \;+\; (T_x - B_x) \frac{C_x^W}{AC_{50,x}^W \;+\; C_x^W} \tag{1}$$

In ToxCast, 1102 chemicals are identified as aromatase inhibitors by the *Tox21-aromatase-inhibition* assay (on MCF-7 human breast cells) with "some reasonable confidence in the estimated parameter values". Among those, 256 chemicals (matching either by CAS number or chemical name) also had exposure estimates in ExpoCast. However, cytotoxicity has been shown to induce many false positive results in ToxCast (Judson et al. 2016). Of the 256 chemicals mentioned above, we kept only the 86 which had an $AC_{90}$ for aromatase inhibition lower than their cytotoxicity $AC_{10}$ (as measured by the ToxCast proliferation decrease assay on T47D human breast cells). The virtual mixtures generated included all of those 86 chemicals.

## *Exposure modeling*

ExpoCast provided the molecular mass, geometric mean, and lower and upper 95% confidence limits of the exposure rate (mg/kg/day) for each chemical present in the randomly generated mixtures. For constant exposure modeling over 10 months, we sampled a rate for each chemical from the corresponding log-normal distribution, but with an SD scaled by the square-root of the number of days of exposure simulated (since a constant exposure rate should be a time average level in that case) and converted it to µmole/kg/minute.

More realistic time-varying exposures over a two-year period were also modeled (similarly to Bertail et al. 2010) using exposure windows of random length and intensity. We first sampled the number *n* of exposure windows for each chemical in the mixture from a scaled exponential distribution (with a rate parameter equal to 5). That yielded on average 145 exposure events over two years (median: 100 events, 1st quartile: 40 events, 3rd quartile: 200 events). The *n* exposures' start and end times were sampled uniformly over the two-year period. The exposure rate during each of the *n* exposure windows was randomly sampled from the log-normal distribution given by ExpoCast, with an SD scaled by the number of days of the exposure window considered (or unscaled if the exposure lasted less than a day).

To get a lower bound on the effect of mixtures, we similarly simulated random non-overlapping exposures to the 86 EDCs selected (*i.e.*, each person was exposed to the 86 chemicals, in random order, at random times, but to only one chemical at a time).

*Pharmacokinetics modeling*

For each chemical in each mixture, a one-compartment PK model was used to estimate its internal concentration (in µM) at steady-state in the case of constant oral exposure or at any point in time in the case of varying oral exposures. Steady-state internal concentrations for chemical *x* were calculated as:

$$C_{x,ss} = \frac{F_x \cdot E_x}{K_{e,x}}, \qquad (2)$$

where $F_x$ is the bioavailability of *x* (unitless), $E_x$ its exposure rate (in µmol/kg/min, sampled as indicated above), and $K_{e,x}$ its total body clearance rate constant (in min$^{-1}$). A body density of 1 was assumed.

For time-varying exposures, internal concentrations were obtained as a function of time by numerical integration of the following differential over a two-year period (with an initial value set to zero):

$$\frac{\partial C_x}{\partial t} = F_x \cdot E_x - K_{e,x} \cdot C_x \qquad (3)$$

We used quantitative structure-activity relationships (QSAR) to obtain central estimates of $F_x$ and $K_{e,x}$ for each of the 86 aromatase inhibitors considered. The robustness of the prediction was evaluated by examining compounds from the training set similar to the target substances, together with literature data and references.

$F_x$ central estimates were obtained at several oral dose levels and were linearly interpolated between dose levels as needed. Beyond the dose rates of 0.001 to 10 mg/day, the $F_x$ value at the closest bound was used).

To take into account the uncertainty affecting the QSAR estimated PK parameters, we randomly sampled $F$ values from Beta distributions (naturally bounded between 0 and 1), with parameters calculated so that the distribution modes corresponded to the interpolated value of $F_x$, with a CV of 20% (for null $F_x$ modes, $a$ and $b$ were set to 1 and 50, yielding a median at 0.01, a 1$^{st}$ quartile at 0.006, and a 3$^{rd}$ quartile at 0.03, approximately). $K_{e,x}$ values were log-normally sampled with a geometric mean equal to the central estimates obtained by QSAR and a geometric SD corresponding to a factor 3.

*Ovarian cycle model*

We adapted the menstrual cycle model presented by Chen and Ward (2013). The model describes the inhibitory and stimulatory effects of hormones E2 and P4 on the hypothalamus-pituitary axis in women (Figure 1). The equations and definitions of all parameters used in the model are given in Methods S2 and Tables S3 and S4 of the Supplemental Material. In the original model, E2 and P4 were assumed to be instantaneously in equilibrium between blood and the ovaries. Instead, we described the kinetics of E2 and P4 by differential equations (eqs. 15-19 in Methods S2). That modification had practically no impact on the time course of the model variables during a normal cycle (equilibrium between blood and ovaries is fast, as assumed by Chen and Ward), but it allowed us to coherently integrate the dynamic aspect of estradiol synthesis inhibition by EDCs. The three additional parameters (blood and ovarian volumes, ovarian blood flow) were obtained from the literature (Table S3).

An additional variable, the ratio of disrupted over basal E2 synthesis rate constants ($ED_{CYP19}$), was introduced to link the internal doses of chemicals in mixtures to aromatase inhibition. $ED_{CYP19}$ was calculated using Hill's model, parametrized with the chemical-specific values provided by ToxCast, as a cumulative product of remaining activity for each of the $m$ chemicals of the mixture considered:

$$ED_{\text{CYP19}} \;=\; \prod_{x=1}^{x=m}\left(1 - \frac{T_x}{100\,(AC_{50,\,x}^{W_i} + C_x^{W_i})}\right) \tag{4}$$

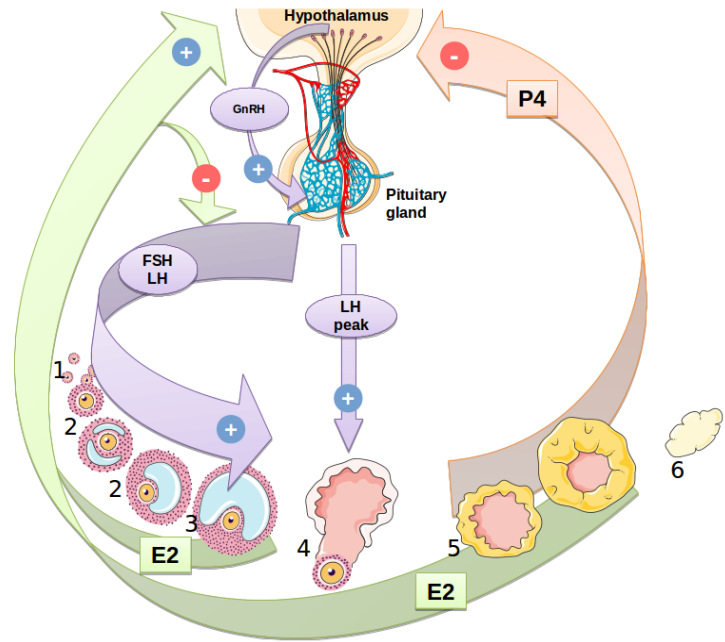

**Figure 1.** Regulatory pathways of the human menstrual cycle, as implemented in the model. During the follicular phase (1: germ cells; 2: developing follicle; 3: mature follicle; 4: ovulation; 5: *corpus luteum* formation; 6: *corpus luteum* degradation), negative feedback by E2 reduces FSH secretion, leading to the selection of one follicle for ovulation. GnRH secretion is promoted by E2 and inhibited by P4, inducing a LH peak and consecutive ovulation. E2 is mainly produced by follicles and *corpus luteum* and P4 by *corpus luteum*.

For constant exposures, $C_x$ was set to the steady-state internal concentrations $C_{x,ss}$. For time-varying exposures, $C_x$ was computed by integration as explained above. The parameters $B_x$ were set to zero because a positive inhibition with no dosage would not make sense. $AC_{50,x}$, $W_x$ and $T_x$ values were randomly sampled using the mean and standard error provided by ToxCast. $AC_{50,x}$ was sampled from a log-normal distribution (its logarithm is actually the ToxCast fitted value). $W_x$ and $T_x$ were sampled from truncated normal distributions. Truncation was from 0 to 10 for $W_x$ (values beyond 10 would be found for some chemicals for which $W_x$ is poorly identified, but have no biological meaning). Truncation was from 0 to 100 for letrozole's $T_x$ (the positive control). For the other chemicals, truncation was from 0 to 10000 over $T_x$ for letrozole to properly rescale the ToxCast $T_x$ values between 0 and 100.

9 / 25

$ED_{CYP19}$ was entered as an input to the ovarian cycle model, which was then solved to obtain the time profile of its output variables over two years of simulated time. In case of constant exposures, we computed the square-root of the sum of the squared Euclidean distances between a reference E2 concentration ($ED_{CYP19}$ set at zero) and the perturbed concentrations (at a fixed set of times) as a summary measure of disruption.

## *Software used*

ACD/Labs Percepta platform modules "ACD/Oral Bioavailability" and "ACD/ PK Explorer" were used for the prediction of oral bioavailability ($F$) and total body clearance rate constant ($K_e$), respectively (see Methods S1, Table S1, Figures S2 and S3 in Supplemental Material). *GNU MCSim* v5.6.5 (www.gnu.org/software/mcsim) (Bois 2009) was used to build the ovarian cycle model. *R* v3.1.1 ([www.R-project.org](www.R-project.org)) (R Development Core Team 2013) with packages *parallel*, *deSolve* and *EnvStats* was used for database processing, numerical integration of the models, and graphics.

## **RESULTS**

## *Estimates of internal dose*

The relationship between constant exposure rates and steady-state internal concentrations for the 86 EDCs considered indicates that exposures ranged from $10^{-8}$ to $10^{-3}$ µmole/kg/ day and the resulting steady-state internal concentrations ranged from $10^{-13}$ to $10^{-3}$ µM (Figure S4, Supplemental Material). The exposure rates and pharmacokinetic parameters were Monte Carlo sampled as explained above. For any single EDC, uncertainty is about a factor of 10 for exposures and a factor of 1000 for the resulting internal concentrations. For time-varying exposures, Figure S5 (Supplemental Material) shows an example of a simulated random two-year time course of internal concentration for lindane. Such profiles were obtained for each chemical in each simulated mixture.

## *Cycle model behavior*

Our implementation of the ovarian cycle model proposed by Chen and Ward (2013) correctly reproduces their results. Human data from McLachlan et al. (1990) and Welt et al. (1999) on LH, FSH, E2 and P4 normal cycles are correctly simulated, except for McLachlan FSH data for which the baseline levels are not well matched. There is a large intra- and inter-variability in hormonal levels across women in those datasets (see Supplemental Material Figure S6). We took the model simulated normal cycling of E2 as the "reference cycle" in the following. Constant exposure scenarios result, at steady-state, in a constant level of aromatase inhibition. In that case, perturbation depends only on that parameter (according to the model assumptions) so the distance between the perturbed and reference cycles is a useful measure of effect (see Figure 2). As aromatase inhibition increases, cycles become increasingly perturbed and exhibit chaotic features (hence the misalignment of the points in Figure 2). At 5% inhibition (95% of normal aromatase activity), cycles are shortened, baseline levels change little, and peak levels either increase or decrease less than proportionally except for LH. Basically, the regulations dampen the effect of perturbation. At about 10% inhibition, LH peaks disappear after about five cycles and a major bifurcation in cycle patterns happens: cycles are further shortened, and baseline levels are much increased (*doubled* for E2 and P4, for example, even though E2 synthesis by aromatase is *decreased*) and peak levels mostly decreased; E2 distance to normal increases up to a maximum (Figure 2). At higher inhibition levels the cycles dampen more and more to disappear completely between 30% and 40% inhibition levels (see Supplemental Material Figures S7-S10). Overall, according to this model, having less than 10% constant inhibition of aromatase activity *in vivo* leads to perturbations of the cycle, which is still under control and should be compatible with normal reproductive function. Beyond 10% inhibition, an actual disruption of the system seems to occur.

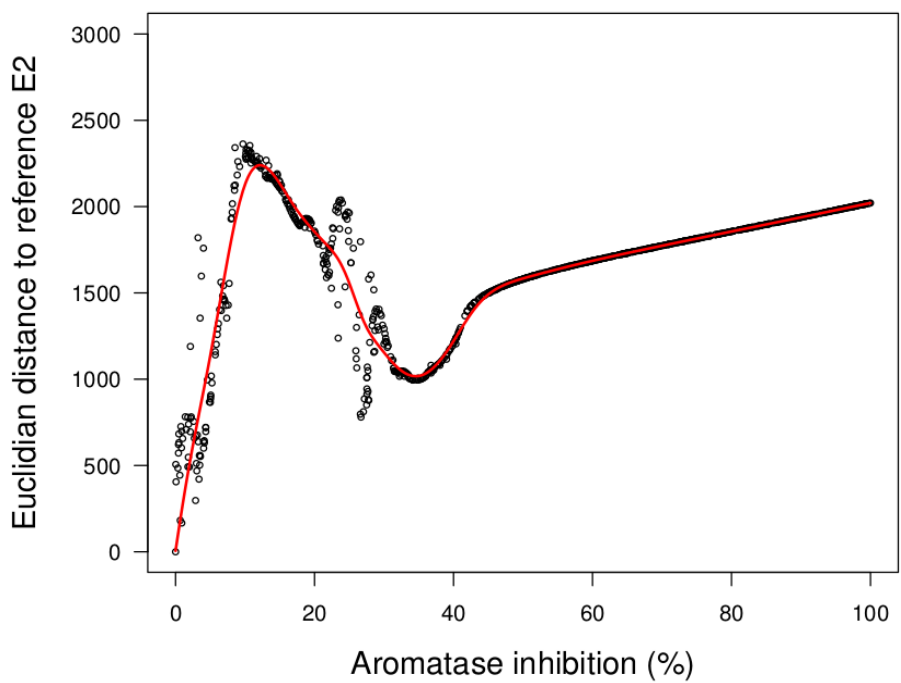

**Figure 2.** Euclidean distance between an estradiol normal cycle and perturbed cycle as a function of aromatase inhibition at steady-state. Distance is computed over two years based on 7301 time points (one every 144 minutes).

## *Effect of single chemicals*

We first simulated a million constant exposures to each of the 86 chemicals considered, taken individually. In that case, despite accounting for uncertainty in exposure levels and dose-response parameters, none of them induced more than 1% aromatase inhibition. Hence, none of them alone was able to induce a significant disruption of the ovarian cycle. Figure 3 places those chemicals on a map with the slope *W* of the Hill dose-response curve at $AC_{50}$, and the log-margin of exposure as coordinates. Margin of exposure was defined as the ratio of the 97.5 percentile of internal concentrations over $AC_{50}$. The $\log_{10}$-margins of the chemicals studied ranged from -10 to -1.8. That means that for all chemicals, the high-end of internal exposure concentrations was at most 1% of $AC_{50}$. In that case, equation 1 shows that the logarithm of aromatase inhibition is approximately the product of *W* times the log-margin of exposure. The color background of the map codes for the resulting "risk index" (*i.e.*, $\log_{10}$-inhibitions) and ranges from less than -2 (1%) to about -80 ($10^{-78}$ %), much too low to elicit changes in ovarian cycles such as seen in Figure 2. Therefore, no effects can be expected from usual exposures to those chemicals when considered alone. Table 1 gives the list of the 10 chemicals for which individual risk is the highest. Note that letrozole is the reference chemical for the Tox21 aromatase inhibition assay, which is consistent with its high rank. The others are found in therapeutic drugs, agrochemicals, food contaminants, consumer products, *etc*.

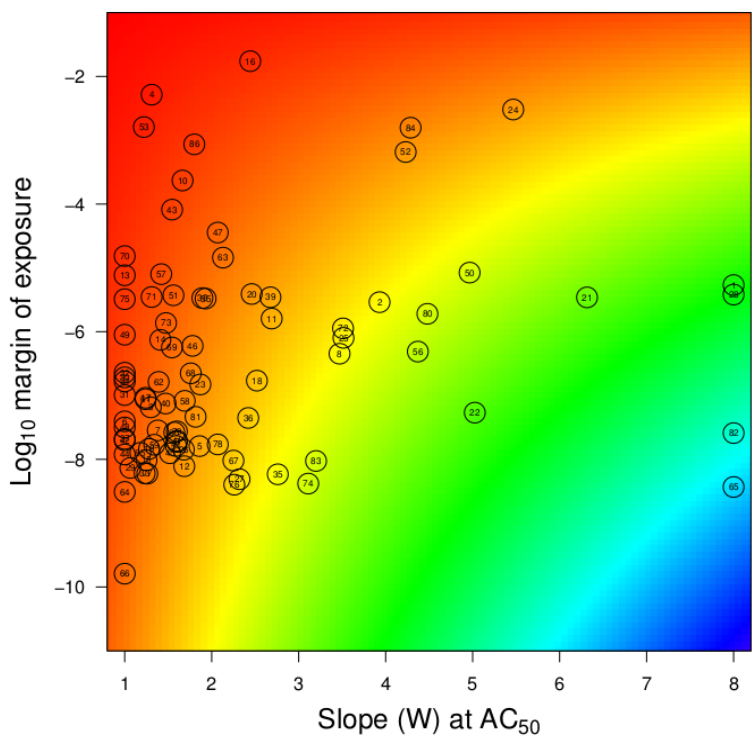

**Figure 3.** Map of the aromatase inhibitors studied over a plane defined by dose-response slope *W* and log-margin of exposure (see text). Colors vary linearly with powers of 10 of aromatase inhibition resulting from *W* and margin of exposure combinations, from red ($10^{-2}$) to blue ($10^{-80}$). For the chemicals-numbers correspondence see Supplemental Material Table S1.

**Table 1**: Top ranking chemicals according to their individual risk indices [a]. Those chemicals are in the top left corner of Figure 3.

| ExpoCast Name | CAS number | Risk Index [a] |
| --- | --- | --- |
| Letrozole | 112809-51-5 | -2.99 |
| Estrone | 53-16-7 | -3.41 |
| Fulvestrant | 129453-61-8 | -4.30 |
| Triflumizole | 68694-11-1 | -4.81 |
| 2,4,7,9-Tetramethyl-5-decyne-4,7-diol | 126-86-3 | -5.11 |
| N-Methyl-2-pyrrolidone | 872-50-4 | -5.49 |
| Rhodamine 6G | 989-38-8 | -5.51 |
| Anastrozole | 120511-73-1 | -6.04 |
| Fenvalerate | 51630-58-1 | -6.05 |
| Imazalil | 35554-44-0 | -6.31 |

[a] Risk Index = $W \times log_{10}(E_{95} / AC_{50})$, where $E_{95}$ is the 95$^{th}$ percentile of the exposure values sampled.

## *Effect of mixtures of chemicals*

We generated one million hypothetical mixtures of the 86 aromatase inhibitors studied and evaluated their global effect on E2 synthesis and resulting ovarian cycle disruption. Figure 4 shows a histogram of the resulting inhibition levels. Depending on the (random) composition of the mixtures, and given the uncertainties on exposures and effect parameters, responses range from zero to 100% inhibition, but on average are very high. Such inhibition levels may lead to perturbation of the ovarian cycle, according to the model. Real-life exposures to chemicals are usually not at constant levels. We simulated time-varying exposures (see Figure S5, Supplemental Material) and the resulting effects on aromatase inhibition and ovarian cycle disruption. To define a lower bound on mixture effects, we first simulated fluctuating EDCs' exposure levels, but without concomitant exposures to them. Interactions can still occur in that case because of storage in the body or because of persistent effects on the ovarian cycle. Figure 5 shows that only 0.3% of the exposures simulated caused more that 10% average aromatase inhibition. The maximum inhibition found was close to 50%. More realistic exposure scenarios do not prohibit concomitant exposures. In that case (Figure 6), the distribution of simulated time-averaged inhibitions is shifted toward higher effects and average inhibitions higher than 20% are not uncommon (yet they do not reach the extreme levels observed on Figure 4).

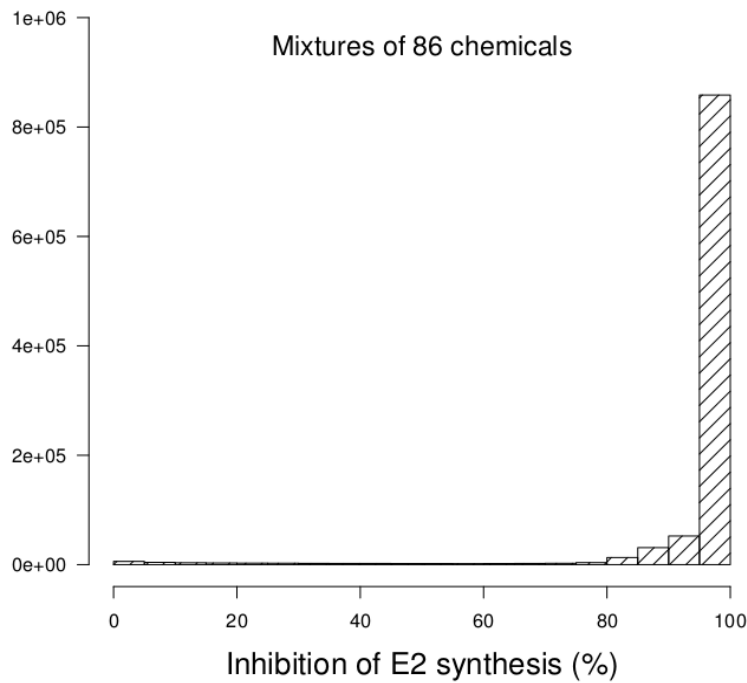

**Figure 4.** Histogram of the percent inhibition of estradiol (E2) synthesis by random mixtures (N = 1 million) of 86 aromatase inhibitors, at constant exposure levels.

Since inhibition changes with time, the distances between normal and perturbed cycles do not follow the pattern of Figure 2 (distances can be much larger) and the link between estradiol inhibition and cycle disruption is harder to establish. Examination of the time-course of the dominant follicle mass ($F$), for 1000 random simulations of mixtures of the 86 chemicals, shows that perturbed cycles typically have a baseline shifted up (which may or may not return to normal) and in irregular succession of peaks (corresponding to ovulation) (Figure 7). The 1000 simulations examined can be classified into four groups. In group 1 (17% of the samples), the cycles are practically normal with no baseline shifts and at most one or two missing ovulations. In group 2 (73% of cases), baseline shifts are always present but without major irregularities in ovulation. Group 3 (7% of cases) has systematic baseline shifts and frequent or prolonged anovulations. Such cycling would clearly impair fertility. In group 4 (3 % of cases), disruption is catastrophic or total. Figure S11 (Supplementary Material) shows the corresponding plots for $E2$ time courses. Judging by them, $E2$ profiles can have a shifted baseline even in normal ovulation profiles, but otherwise the patterns are rather similar.

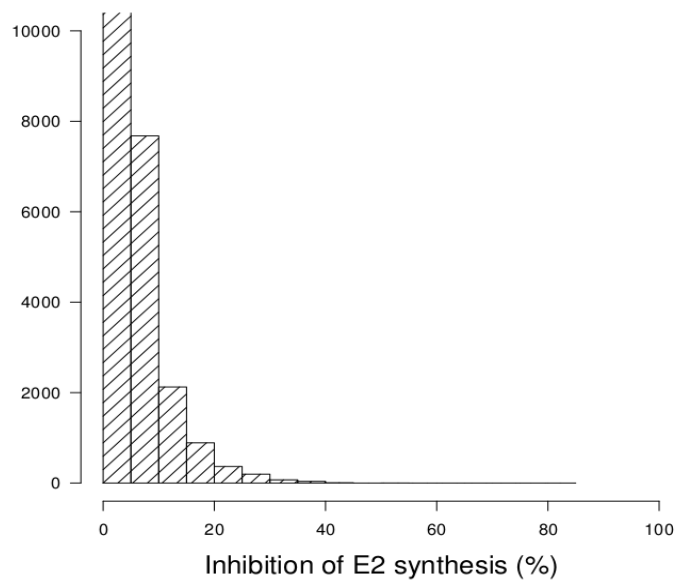

**Figure 5.** Histogram of the percent inhibition of estradiol (E2) synthesis by random mixtures (N = 1 million) of 86 aromatase inhibitors with time-varying non-concomitant exposures. For visibility, the first histogram bar has been truncated (it represents about 990000 simulations).

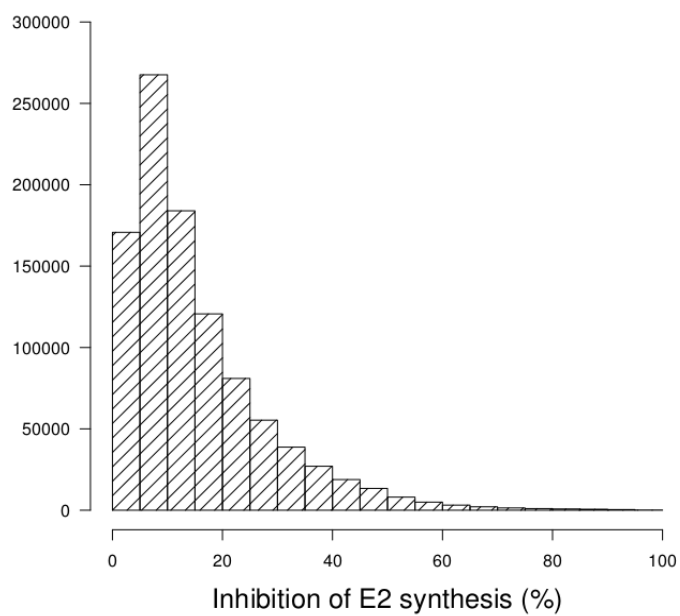

**Figure 6.** Histogram of the percent inhibition of estradiol (E2) synthesis by random mixtures (N = 1 million) of 86 aromatase inhibitors with time-varying exposures.

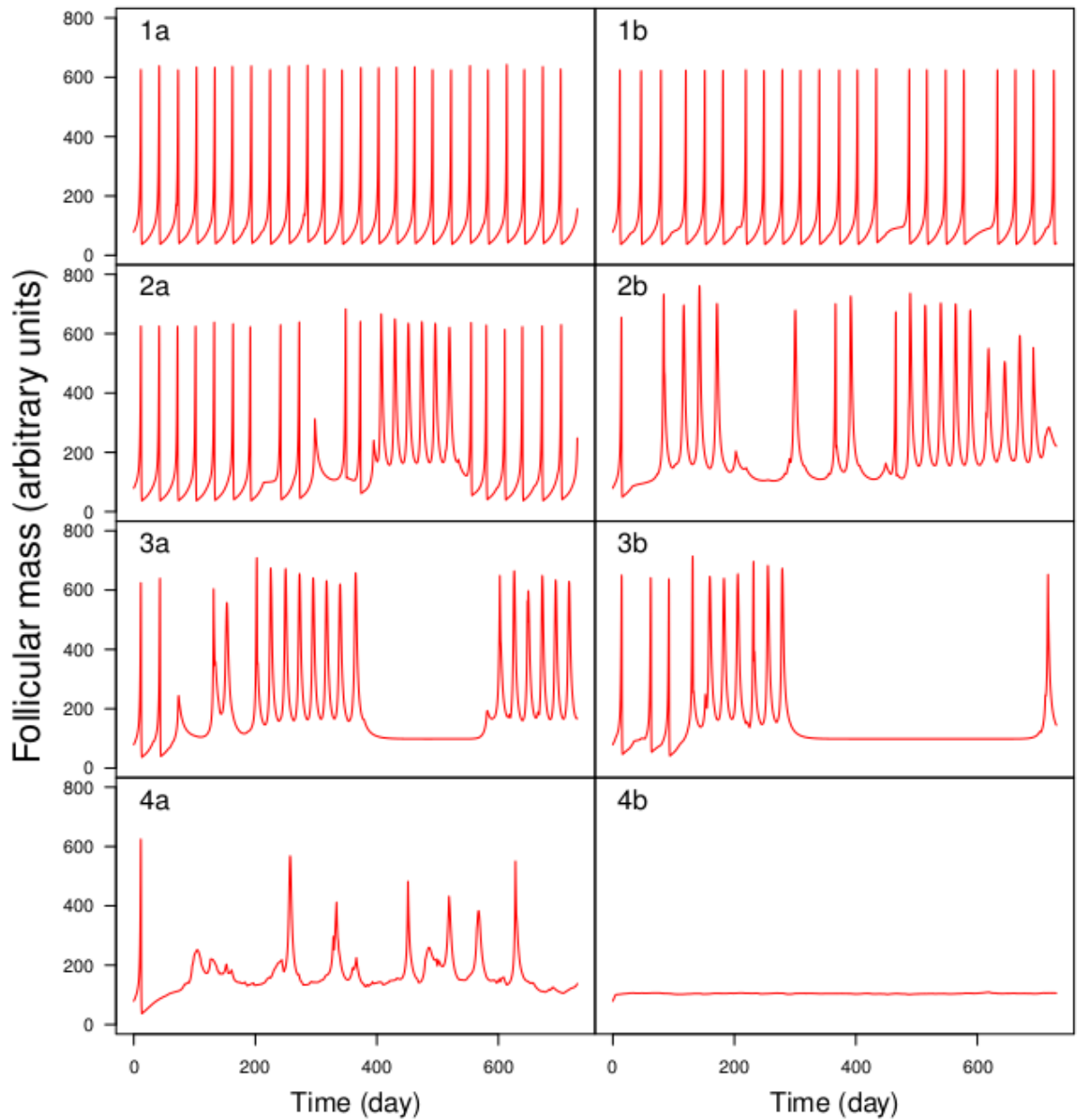

**Figure 7.** Typical simulated time profiles of dominant follicle mass during time-varying exposures to random mixtures of 86 aromatase inhibitors. Four classes (rows) of increasing disruption are illustrated (see text). Left column: least disrupted profile in its class; Right column: most disrupted. Responses range from regular ovulation (panel 1a) to complete disruption (panel 4b).

## DISCUSSION

We linked ToxCast data and ExpoCast estimates of exposures to (mixtures of) aromatase inhibitors, and estimated their effects on the ovarian cycle in women. To our knowledge, this is the first application of computational toxicology and high-throughput testing to assessing the *combined* effect of exposures to a large number of EDCs.

Our approach is predictive and we had to make many assumptions and simplifications. ToxCast and ExpoCast are still incomplete and a full inventory of all the EDCs to which women are exposed is not yet available. So, we were only able to look at a subset of the potential EDCs. We used human exposure estimates from Wambaugh et al. (2013), who give summary statistics for the distribution of exposures to individual chemicals. This allowed us to take the correspondingly large uncertainty into account via Monte Carlo simulations. However, we cannot differentiate between uncertainty and variability in those exposure estimates, and cannot identify subgroups of sensitive individuals. We cannot even focus on women, our target population. More sophisticated exposure models (Isaacs et al. 2014) could help in that respect, but they still deal only with single chemicals and provide no data or estimates about co-exposures. Depending on age, occupation, socioeconomic status, ethnicity, and health condition, we are exposed to different cocktails of chemicals in our diet, workplace, environment, *etc*. (Tornero-Velez et al. 2012). We modeled co-exposures by random sampling, either at a constant level or more realistically with time-varying exposure profiles and hence time-varying mixture complexity. That is still imperfect and we had to make guesses about the distribution of number of exposure windows, for example. We respected the distribution of population exposure levels documented by ExpoCast, but lacking co-exposure information, our estimates might be lower or higher than reality. Efforts are ongoing to collect relevant data; for example, in the European Total Diet Study (Vin et al. 2014). An analysis of such data (Traoré et al. 2016) shows that among 153 synthetic chemicals studied in seven typical French diets (food associations), three of them are aromatase inhibitors according to ToxCast: zearalenone, triadimenol, and lindane. In this regard, mixtures of 86 or more aromatase inhibitors may seem unrealistic, but only food contaminants were studied in Traoré et al. (2016).

We searched for the 86 selected aromatase inhibitors in a database of consumer products marketed in the United States (Gabb and Blake 2016). Briefly, this database was constructed by scraping product information from online retailers. It currently contains 53743 products. Twelve of the 86 aromatase inhibitors were detected in 5701 products (representing 11% of the products in the database). [It is worth noting that none of the 86 aromatase inhibitors are among the volatile fragrance chemicals detected in consumer products (Steinemann 2015; Steinemann et al. 2011), so aromatase inhibitors are unlikely to be hidden in generic "fragrance" or "flavor" designations on consumer product labels.] Two-way combinations of these chemicals were found in 220 products; and the three-way combination of carminic acid, FD&C blue no. 1, and retinol was found in 3 products. This may not seem to point to a large problem but it is an incomplete view of combinatorial exposure. Consider that 3660 out of 4501 makeup products (81%) in the database contain at least one of the aromatase inhibitors considered (carminic acid, retinol, and artificial colors are common ingredients in makeup), and a typical consumer uses several products each day, possibly even several makeup products. This increases the likelihood of combined exposures. Also, no readily available data address associations for all near- and far-field exposures for the 86 aromatase inhibitors, which are used in industrial or agricultural processes (18), consumer product formulations (12), biocidal applications (38), and pharmaceutical drugs (18). These usage categories are likely to be independent, so focusing on a few known associations would only give partial answers and underestimate global risk. We are striving for a more extensive picture. To address the potential overestimation of mixture effects when generating purely random associations, we present the results of a very optimistic exposure scenario (with no co-exposures at all). This gives a lower bound estimate: 0.3% of exposures would lead to more than 10% average aromatase inhibition in women.

ToxCast aromatase inhibition data were obtained by exposing cells *in vitro*. We had no easy way to assess the *in vitro* kinetics of the substances assayed and reconstruction methods (Armitage et al. 2014) require input data that we did not have. We assumed that the nominal assay concentrations were those really experienced by the cells and that equivalent extracellular concentrations *in vivo* would lead to the same levels of aromatase inhibition. That is a typical assumption but it is not necessarily correct (Coecke et al. 2012). To obtain extracellular concentrations *in vivo*, we estimated bioavailability and total clearance with QSAR methods and input them in a simple one-compartment PK model with oral exposure only (even though inhalation or dermal exposures might be more relevant). Again, more sophisticated (PBPK) models and additional PK data would give more precise and accurate predictions (El-Masri et al. 2016; Wambaugh et al. 2015).

On the effect side, we only considered chemicals for which the ToxCast dose-response parameters were estimated with reasonable confidence. This is a conservative choice and additional chemicals would have been included if more relaxed criteria had been chosen (at the price of lower confidence in the results). A concern with large databases such as ToxCast is the quality assurance of the data provided. Aromatase inhibition may not be the most sensitive toxicity endpoint, for example, or it may be due to a burst effect of cytotoxicity (Judson et al. 2016). From the original set of 256 aromatase inhibitors for which we had exposure estimates and ToxCast data, 170 were screened off based on cytotoxicity. Those were mostly weak inhibitors (data not shown). For the screen, we kept only the 86 substances which had an aromatase inhibition $AC_{90}$ lower than their cytotoxicity $AC_{10}$ – measured by the ToxCast proliferation decrease assay on T47D human breast cells, a cell type similar to the MCF-7 used by the aromatase inhibition assays (Aka and Lin 2012). We preferred that screen to the omnibus z-score criterion, which aggregates cytotoxicity results from different cell types and species. For the remaining 86 substances, we may still have downplayed other types of toxicity. Cytotoxicity, by the way, is not a "negligible" endpoint, and an evaluation of the cytotoxicity of mixtures would be interesting in its own right.

The evidence provided by ToxCast is also not perfectly predictive of human *in vivo*. For example, the Tox21 aromatase inhibition assay uses the MCF-7 breast cancer cell lines which might not respond as normal ovary cells. In addition, our model of the ovarian cycle (Chen and Ward 2013) describes only approximately the complex dynamic interactions between ovarian follicular growth and hormonal homeostasis. The hypothalamus and pituitary gland are treated as a single compartment and the description of the central hormonal controls is simplified. More complex models have been proposed (Hendrix et al. 2014) but they still make many assumptions and do not seem to offer dramatically better performance. Note also that we treated the parameters of the ovarian cycle as constant, when in fact they are affected by both uncertainty and variability in response to EDCs, adding to the tails of the distributions of our results. However, we did not have sufficient information to define statistical distributions for those parameters.

In terms of results, an obvious question is that of the "bad" actors, *i.e.*, the chemicals responsible for the predicted effects. The answer is provided by Figure 3 (and partially by the top-ten Table 1), because here the impact of individual chemicals on aromatase is only conditioned by internal dose and inhibition potency. Figure 3 is actually a useful prioritization tool. It is rather simple to construct and does not require running the ovarian cycle (the PK model is needed). However, it does not give an answer in terms of magnitude of effect at the subject level. For that we need the whole body ovarian cycle model.

One of the consistent features of E2 cycle perturbation that we found is that the baseline (inter-ovulation) levels of E2 tend to *increase* (to about twice the normal level) in response to aromatase inhibition (which implies a lower rate of E2 synthesis by aromatase). That counterintuitive feature of the complex cycle dynamic is induced by the CNS feedback. Passed a certain level of inhibition, the control of E2 still "works" (peaks are still observed) but puts the baseline at a higher value. We do not have confirmation that this is the case in women exposed to EDCs, but that would be interesting information and a potential biomarker of effects. Note also that we lack good measures of perturbation for such complex systems. We used visual inspection to classify cycles for 1000 time-varying exposures (Figure 7). Perturbation analysis of more cycles would require more sophisticated tools. Finally, many other perturbation pathways exist for ovarian cycle disruption which were not accounted for (*e.g.*, actions mediated by the androgen or estrogen receptors). The simplicity of our model also precludes investigation of synergistic or antagonistic effects that could be due to metabolic or toxicodynamic interactions (Cheng and Bois 2011).

# CONCLUSIONS

High-throughput data collection requires high-throughput analysis, extrapolation, and decision tools if we want to avoid a bottleneck and accumulation of unused data. We developed such a tool, making use of our increasing understanding of toxicity mechanisms.

This exercise in prediction points to large data gaps. Our knowledge of exposure to actual mixtures is minimal, except in a few cases (tar, tobacco smoke, *etc.*) For the chemicals studied here, quantitative knowledge of their routes of exposure and PK parameters is also lacking. Our knowledge of endocrine disruption mechanisms is still in its infancy. We should probably check, for example, the inhibition potential of the 86 substances studied here with better tests and a better characterization of the *in vitro* fate of the chemicals. At least we now have a prioritized list and some reason to deepen our investigations. We should also research relevant human biomarkers of exposure or effects for validation of the results and better actions.

The basic assumption of regulatory practice and most risk assessments is that keeping individual chemicals under control with reasonable safety factors will keep their joint effects at bay. That may not be the case. Some women have various levels of ovarian dysfunction that can be caused by various internal and external factors. Environmental chemical exposures add onto that background (National Research Council 2009; Zeise et al. 2013). We found that even though individual chemicals are probably "safe" as used now, their joint effects, when they exceed a few dozens in number, can lead to severe disruption of the ovarian cycle in women in a sizable number of cases. Obviously, this is no definite proof that some fertility problems can be caused by real-life exposures to EDCs. Still, risk assessment practice and regulations should start thinking of the problem of mixtures not as an unsolvable one, but as needing a clearly laid out research agenda. In the case in point, the simple graphical map of internal dose and potency we propose could already be used for prioritization. Given the magnitude range of the predicted effects on ovarian cycling, while waiting for confirmation of our results, a cautionary attitude should be adopted.

**Supplemental Material**

**A high-throughput analysis of ovarian cycle disruption by mixtures of aromatase inhibitors**

Frederic Y. Bois, Nazanin Golbamaki-Bakhtyari, Simona Kovarich, Cleo Tebby, Henry A. Gabb,

Emmanuel Lemazurier

**Table of contents**

**Figure S1:** Computational workflow.

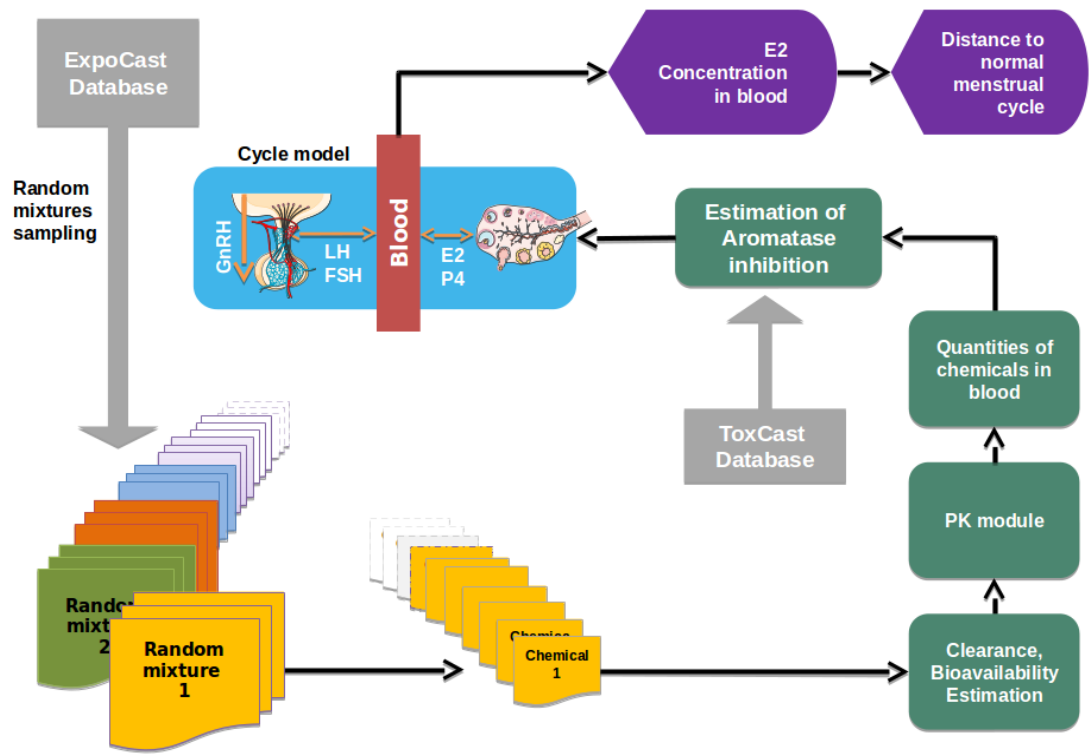

**Figure S1.** Computational workflow. Each box represents a task to be executed or a result obtained and arrows represent either data flow or execution dependencies between different tasks. Random mixtures of chemicals are sampled from ExpoCast. The kinetics of each chemical in the human body are computed using a one-compartment PK model. The resulting aromatase inhibition is computed from ToxCast dose-response information. The impact of that inhibition on the menstrual cycle is estimated using an ovarian cycle model adapted from Chen and Ward (2013).

**Methods S1:** PK modules of ACD/Labs.

**The ACD/Oral Bioavailability module** uses a combination of probabilistic and mechanistic models to predict oral bioavailability based on a dataset of oral bioavailability data for 788 substances. The module predicts the fraction of the specified dose that reaches systemic circulation after oral administration (*F*), with the possibility to explore dose-dependence of *F*. The default oral dose used in calculations is 50 mg. The mathematical model used for simulations in the Oral Bioavailability (and PK Explorer) modules is based on differential equations that consider solubility in the gastrointestinal tract, passive-absorption in jejunum, elimination (total body clearance), and volume of distribution. Simulations performed with the Oral Bioavailability module ignore first-pass metabolism in liver and gut. Quantitative *F* values are calculated as a ratio of AUCs after oral and intravenous administration. A number of endpoints affecting oral bioavailability are also predicted, including solubility (dose/solubility ratio), stability in acidic media, intestinal membrane permeability by passive or active transport (with a summary of transporters where relevant), P-gp efflux (multi-drug transporter P-glycoprotein) and first pass metabolism in the liver. For the majority of these parameters, ACD/Labs Percepta allows the user to assesses the confidence in the predicted result by means of reliability index and/or displaying similar structures from the training set along with their experimental test results. Since the bioavailability predictions are based on a mechanistic simulation model rather that statistical fitting to data, the validation procedure was not based on a conventional training/test set approach. Instead, a set of clinical fraction absorbed ($f_a$) data together with dosage information for 28 drugs reported by Parrot and Lavé (2002) was used for validation purposes. The model performance was assessed qualitatively, by evaluating the accuracy of three-class classification (i.e., "low" for $f_a \leq 33\%$, "moderate" for $33\% < f_a < 66\%$, "high" for $f_a \geq 66\%$), and quantitatively, using the Residual Mean Square Error statistic and visual inspection of the correlation between observed and predicted $f_a$ values for the drugs considered. For the majority of drugs, calculated values were in good agreement with clinical data, resulting in 80% correct classification and an RMSE equal to 17%.

**ACD/PK Explorer** is a tool for exploring pharmacokinetic relationships (*e.g.*, *F*–Dose relationship) using physico-chemical parameters as inputs for calculations. Total body clearance rate constant (*Ke*) is among the required parameters of our PK model. *Ke* was estimated using a set of fragmental ionization class specific models. Hydrophobicity (logP) and ionization (pKa) are the principal descriptors for the prediction of *Ke*. These descriptors were calculated *in silico*. Technical details on those models and equations are not available since they are proprietary information of ACD/Labs.

**Table S1:** Predictions of bioavailability (*F*) and total body clearance rate constant ($K_e$). The ACD/Labs Percepta Oral Bioavailability module was used for predictions of oral F values. F predictions depend on oral dose and are given for an array of doses. The module ACD/ PK Explorer was used for the prediction of $K_e$ values.

| ID | CAS | F (0.001 mg) | F (0.01 mg) | F (0.1 mg) | F (1 mg) | F (5 mg) | F (10 mg) | F (50 mg) | F (100 mg) | F (500 mg) | F (1000 mg) | $K_e$ (min$^{-1}$) |
|---|---|---|---|---|---|---|---|---|---|---|---|---|
| 1 | 101-20-2 | 0.2805 | 0.2804 | 0.2798 | 0.2739 | 0.2505 | 0.2269 | 0.1342 | 0.0919 | 0.0295 | 0.0168 | 0.00083 |
| 2 | 101-80-4 | 0.9760 | 0.9760 | 0.9760 | 0.9759 | 0.9754 | 0.9748 | 0.9661 | 0.9486 | 0.5863 | 0.3625 | 0.002 |
| 3 | 112281-77-3 | 0.9907 | 0.9907 | 0.9907 | 0.9907 | 0.9903 | 0.9899 | 0.9826 | 0.9575 | 0.5201 | 0.3202 | 0.0011 |
| 4 | 112809-51-5 | 0.9919 | 0.9919 | 0.9919 | 0.9919 | 0.9917 | 0.9915 | 0.9880 | 0.9771 | 0.6065 | 0.3808 | 0.00041 |
| 5 | 114369-43-6 | 0.7271 | 0.7270 | 0.7261 | 0.7175 | 0.6807 | 0.6385 | 0.4214 | 0.2983 | 0.0986 | 0.0566 | 0.0011 |
| 6 | 115-39-9 | 0.2164 | 0.2164 | 0.2158 | 0.2104 | 0.1897 | 0.1694 | 0.0956 | 0.0644 | 0.0201 | 0.0114 | 0.00087 |
| 7 | 116255-48-2 | 0.9884 | 0.9884 | 0.9884 | 0.9882 | 0.9872 | 0.9856 | 0.9580 | 0.8774 | 0.3949 | 0.2374 | 0.0011 |
| 8 | 117-39-5 | 0.9533 | 0.9533 | 0.9533 | 0.9532 | 0.9528 | 0.9521 | 0.9429 | 0.9247 | 0.5982 | 0.3723 | 0.0062 |
| 9 | 1192-52-5 | 0.9942 | 0.9942 | 0.9942 | 0.9942 | 0.9942 | 0.9942 | 0.9942 | 0.9942 | 0.9938 | 0.9933 | 0.0015 |
| 10 | 120511-73-1 | 0.9916 | 0.9916 | 0.9916 | 0.9916 | 0.9915 | 0.9915 | 0.9910 | 0.9901 | 0.9455 | 0.7218 | 0.0009 |
| 11 | 125-84-8 | 0.9936 | 0.9936 | 0.9936 | 0.9936 | 0.9936 | 0.9936 | 0.9935 | 0.9932 | 0.9899 | 0.9812 | 0.0019 |
| 12 | 126-72-7 | 0.9908 | 0.9908 | 0.9908 | 0.9907 | 0.9904 | 0.9899 | 0.9826 | 0.9586 | 0.5198 | 0.3193 | 0.0012 |
| 13 | 126-86-3 | 0.9938 | 0.9938 | 0.9938 | 0.9938 | 0.9937 | 0.9935 | 0.9915 | 0.9862 | 0.6964 | 0.4480 | 0.0017 |
| 14 | 1260-17-9 | 0.0078 | 0.0078 | 0.0078 | 0.0078 | 0.0078 | 0.0078 | 0.0078 | 0.0078 | 0.0078 | 0.0078 | 0.0049 |
| 15 | 129-17-9 | 0.0000 | 0.0000 | 0.0000 | 0.0000 | 0.0000 | 0.0000 | 0.0000 | 0.0000 | 0.0000 | 0.0000 | 0.0071 |
| 16 | 129453-61-8 | 0.0067 | 0.0067 | 0.0066 | 0.0064 | 0.0056 | 0.0049 | 0.0026 | 0.0017 | 0.0005 | 0.0003 | 0.0002 |
| 17 | 131341-86-1 | 0.9393 | 0.9393 | 0.9390 | 0.9360 | 0.9223 | 0.9039 | 0.7408 | 0.5798 | 0.2155 | 0.1264 | 0.0002 |
| 18 | 13171-00-1 | 0.9836 | 0.9836 | 0.9835 | 0.9830 | 0.9802 | 0.9759 | 0.9089 | 0.7791 | 0.3177 | 0.1886 | 0.0012 |
| 19 | 131860-33-8 | 0.9603 | 0.9603 | 0.9601 | 0.9581 | 0.9481 | 0.9337 | 0.7802 | 0.6097 | 0.2225 | 0.1298 | 0.00068 |

| ID | CAS | F (0.001 mg) | F (0.01 mg) | F (0.1 mg) | F (1 mg) | F (5 mg) | F (10 mg) | F (50 mg) | F (100 mg) | F (500 mg) | F (1000 mg) | $K_e$ (min$^{-1}$) |
|---|---|---|---|---|---|---|---|---|---|---|---|---|
| 20 | 13292-46-1 | 0.1755 | 0.1755 | 0.1755 | 0.1755 | 0.1755 | 0.1755 | 0.1755 | 0.1755 | 0.1754 | 0.1713 | 0.0019 |
| 21 | 133-06-2 | 0.8723 | 0.8723 | 0.8717 | 0.8665 | 0.8427 | 0.8128 | 0.6083 | 0.4543 | 0.1600 | 0.0930 | 0.0017 |
| 22 | 133-07-3 | 0.5715 | 0.5714 | 0.5705 | 0.5623 | 0.5282 | 0.4910 | 0.3167 | 0.2236 | 0.0742 | 0.0427 | 0.0015 |
| 23 | 133855-98-8 | 0.9064 | 0.9064 | 0.9059 | 0.9016 | 0.8814 | 0.8551 | 0.6557 | 0.4932 | 0.1743 | 0.1012 | 0.00075 |
| 24 | 137-26-8 | 0.9764 | 0.9764 | 0.9763 | 0.9750 | 0.9690 | 0.9603 | 0.8526 | 0.7011 | 0.2741 | 0.1619 | 0.0022 |
| 25 | 137-30-4 | 0.8689 | 0.8689 | 0.8689 | 0.8689 | 0.8689 | 0.8689 | 0.8689 | 0.8689 | 0.8689 | 0.8689 | 0.0032 |
| 26 | 14233-37-5 | 0.8501 | 0.8500 | 0.8494 | 0.8430 | 0.8143 | 0.7789 | 0.5556 | 0.4044 | 0.1376 | 0.0794 | 0.0014 |
| 27 | 15299-99-7 | 0.9827 | 0.9827 | 0.9826 | 0.9819 | 0.9785 | 0.9734 | 0.8969 | 0.7597 | 0.3053 | 0.1809 | 0.0014 |
| 28 | 155990-20-8 | 0.9920 | 0.9920 | 0.9920 | 0.9920 | 0.9920 | 0.9920 | 0.9920 | 0.9920 | 0.9919 | 0.9915 | 0.0013 |
| 29 | 15972-60-8 | 0.9952 | 0.9952 | 0.9952 | 0.9952 | 0.9952 | 0.9952 | 0.9950 | 0.9948 | 0.9901 | 0.9624 | 0.0019 |
| 30 | 16423-68-0 | 0.0002 | 0.0002 | 0.0002 | 0.0001 | 0.0001 | 0.0001 | 0.0001 | 0.0000 | 0.0000 | 0.0000 | 0.00065 |
| 31 | 175013-18-0 | 0.9892 | 0.9892 | 0.9892 | 0.9890 | 0.9882 | 0.9870 | 0.9656 | 0.8972 | 0.4116 | 0.2477 | 0.0012 |
| 32 | 17924-92-4 | 0.9958 | 0.9958 | 0.9958 | 0.9958 | 0.9958 | 0.9958 | 0.9953 | 0.9943 | 0.9495 | 0.7248 | 0.0022 |
| 33 | 1918-02-1 | 66.39 | 66.39 | 66.39 | 66.39 | 66.39 | 66.39 | 66.39 | 66.39 | 66.39 | 66.39 | 0.0025 |
| 34 | 199171-88-5 | 0.9916 | 0.9916 | 0.9916 | 0.9916 | 0.9916 | 0.9916 | 0.9913 | 0.9909 | 0.9811 | 0.9148 | 0.00099 |
| 35 | 2310-17-0 | 0.7593 | 0.7592 | 0.7584 | 0.7503 | 0.7153 | 0.6745 | 0.4547 | 0.3245 | 0.1083 | 0.0623 | 0.0014 |
| 36 | 2312-35-8 | 0.8301 | 0.8301 | 0.8294 | 0.8223 | 0.7913 | 0.7535 | 0.5266 | 0.3800 | 0.1280 | 0.0737 | 0.0013 |
| 37 | 2353-45-9 | 0.0000 | 0.0000 | 0.0000 | 0.0000 | 0.0000 | 0.0000 | 0.0000 | 0.0000 | 0.0000 | 0.0000 | 0.0047 |
| 38 | 24169-02-6 | 0.6276 | 0.6275 | 0.6266 | 0.6173 | 0.5789 | 0.5366 | 0.3397 | 0.2372 | 0.0773 | 0.0442 | 0.00089 |
| 39 | 2425-06-1 | 0.5641 | 0.5641 | 0.5632 | 0.5544 | 0.5186 | 0.4797 | 0.3031 | 0.2121 | 0.0695 | 0.0398 | 0.0016 |
| 40 | 29091-21-2 | 0.2182 | 0.2182 | 0.2177 | 0.2127 | 0.1932 | 0.1739 | 0.1007 | 0.0685 | 0.0218 | 0.0124 | 0.00054 |
| 41 | 30399-84-9 | 0.9043 | 0.9042 | 0.9037 | 0.8990 | 0.8772 | 0.8488 | 0.6383 | 0.4740 | 0.1645 | 0.0952 | 0.0011 |

| ID | CAS | F (0.001 mg) | F (0.01 mg) | F (0.1 mg) | F (1 mg) | F (5 mg) | F (10 mg) | F (50 mg) | F (100 mg) | F (500 mg) | F (1000 mg) | $K_e$ (min$^{-1}$) |
|---|---|---|---|---|---|---|---|---|---|---|---|---|
| 42 | 311-45-5 | 0.9965 | 0.9965 | 0.9965 | 0.9965 | 0.9965 | 0.9965 | 0.9965 | 0.9964 | 0.9957 | 0.9940 | 0.0023 |
| 43 | 35554-44-0 | 0.9924 | 0.9924 | 0.9924 | 0.9924 | 0.9924 | 0.9923 | 0.9916 | 0.9903 | 0.8887 | 0.6276 | 0.0012 |
| 44 | 3844-45-9 | 0.0000 | 0.0000 | 0.0000 | 0.0000 | 0.0000 | 0.0000 | 0.0000 | 0.0000 | 0.0000 | 0.0000 | 0.007 |
| 45 | 43121-43-3 | 0.9919 | 0.9919 | 0.9919 | 0.9918 | 0.9917 | 0.9915 | 0.9889 | 0.9819 | 0.6581 | 0.4188 | 0.0012 |
| 46 | 49562-28-9 | 0.9326 | 0.9326 | 0.9322 | 0.9286 | 0.9116 | 0.8886 | 0.6955 | 0.5259 | 0.1858 | 0.1079 | 0.0016 |
| 47 | 50-41-9 | 0.9197 | 0.9197 | 0.9193 | 0.9151 | 0.8954 | 0.8692 | 0.6612 | 0.4919 | 0.1704 | 0.0986 | 0.00014 |
| 48 | 50594-66-6 | 0.6879 | 0.6878 | 0.6868 | 0.6771 | 0.6361 | 0.5904 | 0.3727 | 0.2590 | 0.0836 | 0.0477 | 0.00064 |
| 49 | 51630-58-1 | 0.0101 | 0.0101 | 0.0100 | 0.0098 | 0.0087 | 0.0077 | 0.0042 | 0.0028 | 0.0009 | 0.0005 | 0.0014 |
| 50 | 52-01-7 | 0.9256 | 0.9255 | 0.9251 | 0.9209 | 0.9011 | 0.8746 | 0.6651 | 0.4948 | 0.1714 | 0.0991 | 0.0027 |
| 51 | 520-18-3 | 0.9892 | 0.9892 | 0.9892 | 0.9892 | 0.9890 | 0.9888 | 0.9865 | 0.9819 | 0.8086 | 0.5408 | 0.0015 |
| 52 | 521-18-6 | 0.9647 | 0.9647 | 0.9645 | 0.9624 | 0.9521 | 0.9375 | 0.7852 | 0.6166 | 0.2272 | 0.1328 | 0.0036 |
| 53 | 53-16-7 | 0.9600 | 0.9600 | 0.9598 | 0.9578 | 0.9480 | 0.9342 | 0.7886 | 0.6236 | 0.2321 | 0.1360 | 0.0012 |
| 54 | 55219-65-3 | 0.9917 | 0.9917 | 0.9917 | 0.9917 | 0.9914 | 0.9910 | 0.9858 | 0.9684 | 0.5578 | 0.3462 | 0.0013 |
| 55 | 57-63-6 | 0.8500 | 0.8500 | 0.8493 | 0.8430 | 0.8149 | 0.7801 | 0.5593 | 0.4083 | 0.1395 | 0.0806 | 0.0011 |
| 56 | 577-11-7 | 0.8287 | 0.8286 | 0.8278 | 0.8198 | 0.7841 | 0.7411 | 0.4968 | 0.3509 | 0.1149 | 0.0658 | 0.0055 |
| 57 | 58-14-0 | 0.9688 | 0.9687 | 0.9686 | 0.9672 | 0.9600 | 0.9488 | 0.7915 | 0.6000 | 0.2057 | 0.1184 | 0.0014 |
| 58 | 58-89-9 | 0.9227 | 0.9227 | 0.9223 | 0.9188 | 0.9024 | 0.8808 | 0.7044 | 0.5441 | 0.1994 | 0.1167 | 0.0018 |
| 59 | 60168-88-9 | 0.8674 | 0.8673 | 0.8667 | 0.8610 | 0.8352 | 0.8027 | 0.5871 | 0.4324 | 0.1493 | 0.0864 | 0.0011 |
| 60 | 60-57-1 | 0.3142 | 0.3142 | 0.3135 | 0.3071 | 0.2819 | 0.2562 | 0.1532 | 0.1054 | 0.0339 | 0.0194 | 0.0012 |
| 61 | 62924-70-3 | 0.3046 | 0.3045 | 0.3038 | 0.2971 | 0.2707 | 0.2442 | 0.1422 | 0.0968 | 0.0307 | 0.0175 | 0.00024 |
| 62 | 633-96-5 | 0.7686 | 0.7686 | 0.7686 | 0.7686 | 0.7686 | 0.7686 | 0.7684 | 0.7681 | 0.7457 | 0.6995 | 0.0027 |

| ID | CAS | F (0.001 mg) | F (0.01 mg) | F (0.1 mg) | F (1 mg) | F (5 mg) | F (10 mg) | F (50 mg) | F (100 mg) | F (500 mg) | F (1000 mg) | $K_e$ (min$^{-1}$) |
|---|---|---|---|---|---|---|---|---|---|---|---|---|
| 63 | 643-79-8 | 0.9938 | 0.9938 | 0.9938 | 0.9938 | 0.9938 | 0.9938 | 0.9938 | 0.9938 | 0.9934 | 0.9929 | 0.0014 |
| 64 | 6459-94-5 | 0.0000 | 0.0000 | 0.0000 | 0.0000 | 0.0000 | 0.0000 | 0.0000 | 0.0000 | 0.0000 | 0.0000 | 0.00093 |
| 65 | 6610-29-3 | 0.4977 | 0.4977 | 0.4977 | 0.4977 | 0.4977 | 0.4977 | 0.4977 | 0.4977 | 0.4953 | 0.4840 | 0.0024 |
| 66 | 6625-46-3 | 0.0001 | 0.0001 | 0.0001 | 0.0001 | 0.0001 | 0.0001 | 0.0001 | 0.0001 | 0.0001 | 0.0001 | 0.0034 |
| 67 | 67-20-9 | 0.9912 | 0.9912 | 0.9912 | 0.9912 | 0.9911 | 0.9910 | 0.9898 | 0.9875 | 0.9150 | 0.6701 | 0.0016 |
| 68 | 67747-09-5 | 0.9904 | 0.9904 | 0.9903 | 0.9902 | 0.9896 | 0.9888 | 0.9742 | 0.9245 | 0.4481 | 0.2716 | 0.0013 |
| 69 | 68-26-8 | 0.9807 | 0.9807 | 0.9806 | 0.9799 | 0.9762 | 0.9706 | 0.8870 | 0.7427 | 0.2928 | 0.1729 | 0.0011 |
| 70 | 68694-11-1 | 0.9918 | 0.9918 | 0.9918 | 0.9918 | 0.9917 | 0.9916 | 0.9899 | 0.9856 | 0.7227 | 0.4689 | 0.00044 |
| 71 | 79241-46-6 | 0.9150 | 0.9150 | 0.9145 | 0.9103 | 0.8906 | 0.8646 | 0.6624 | 0.4965 | 0.1742 | 0.1010 | 0.0011 |
| 72 | 84-61-7 | 0.9670 | 0.9670 | 0.9668 | 0.9648 | 0.9553 | 0.9415 | 0.7935 | 0.6247 | 0.2304 | 0.1347 | 0.0029 |
| 73 | 85509-19-9 | 0.9813 | 0.9813 | 0.9812 | 0.9805 | 0.9771 | 0.9718 | 0.8933 | 0.7539 | 0.3011 | 0.1782 | 0.00051 |
| 74 | 86-50-0 | 0.9902 | 0.9902 | 0.9902 | 0.9899 | 0.9883 | 0.9860 | 0.9459 | 0.8454 | 0.3659 | 0.2190 | 0.002 |
| 75 | 872-50-4 | 0.9959 | 0.9959 | 0.9959 | 0.9959 | 0.9959 | 0.9959 | 0.9959 | 0.9959 | 0.9959 | 0.9959 | 0.0022 |
| 76 | 87392-12-9 | 0.9951 | 0.9951 | 0.9951 | 0.9951 | 0.9951 | 0.9951 | 0.9948 | 0.9944 | 0.9835 | 0.8928 | 0.0019 |
| 77 | 87674-68-8 | 0.9960 | 0.9960 | 0.9960 | 0.9960 | 0.9960 | 0.9960 | 0.9959 | 0.9959 | 0.9952 | 0.9936 | 0.0021 |
| 78 | 88-58-4 | 0.9925 | 0.9925 | 0.9925 | 0.9924 | 0.9924 | 0.9923 | 0.9916 | 0.9901 | 0.8756 | 0.6122 | 0.0012 |
| 79 | 88671-89-0 | 0.9916 | 0.9916 | 0.9916 | 0.9914 | 0.9905 | 0.9892 | 0.9648 | 0.8909 | 0.4072 | 0.2453 | 0.0019 |
| 80 | 88-72-2 | 0.9955 | 0.9955 | 0.9955 | 0.9955 | 0.9955 | 0.9955 | 0.9954 | 0.9952 | 0.9906 | 0.9465 | 0.0019 |
| 81 | 88-85-7 | 0.9950 | 0.9950 | 0.9950 | 0.9950 | 0.9950 | 0.9950 | 0.9950 | 0.9950 | 0.9950 | 0.9950 | 0.002 |
| 82 | 941-69-5 | 0.9955 | 0.9955 | 0.9955 | 0.9955 | 0.9955 | 0.9955 | 0.9955 | 0.9954 | 0.9950 | 0.9943 | 0.0019 |
| 83 | 94361-06-5 | 0.9921 | 0.9921 | 0.9921 | 0.9921 | 0.9920 | 0.9918 | 0.9901 | 0.9861 | 0.7296 | 0.4747 | 0.0019 |

| ID | CAS | F (0.001 mg) | F (0.01 mg) | F (0.1 mg) | F (1 mg) | F (5 mg) | F (10 mg) | F (50 mg) | F (100 mg) | F (500 mg) | F (1000 mg) | $K_e$ (min$^{-1}$) |
|---|---|---|---|---|---|---|---|---|---|---|---|---|
| 84 | 97-77-8 | 0.8355 | 0.8354 | 0.8347 | 0.8281 | 0.7985 | 0.7624 | 0.5412 | 0.3939 | 0.1343 | 0.0775 | 0.0024 |
| 85 | 98-51-1 | 0.9778 | 0.9778 | 0.9777 | 0.9768 | 0.9724 | 0.9661 | 0.8818 | 0.7463 | 0.3045 | 0.1811 | 0.0024 |
| 86 | 989-38-8 | 0.9518 | 0.9518 | 0.9515 | 0.9488 | 0.9359 | 0.9175 | 0.7391 | 0.5641 | 0.2002 | 0.1162 | 0.00041 |

Figure S2: Bioavailabilities as predicted by the ACD/Labs software for each of 86 EDC.

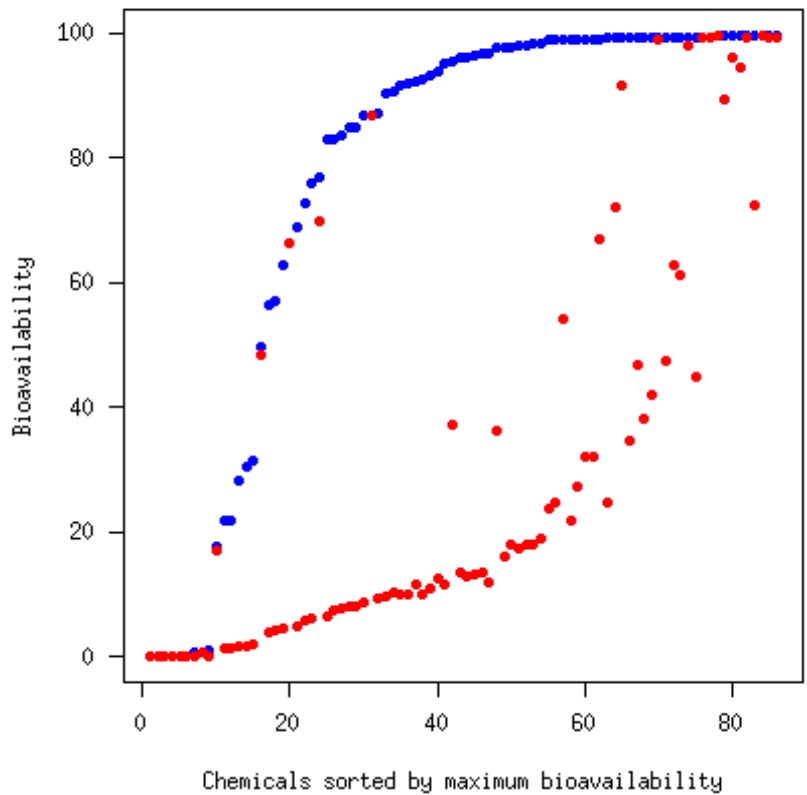

**Figure S2:** Bioavailability as predicted by ACD/Labs for each of 86 chemicals. The blue and red dots show respectively the bioavailability for oral administration of 0.001 mg and 1000 mg of each chemical.

Figure S3: Histogram of the half-life of each of the 86 EDC.

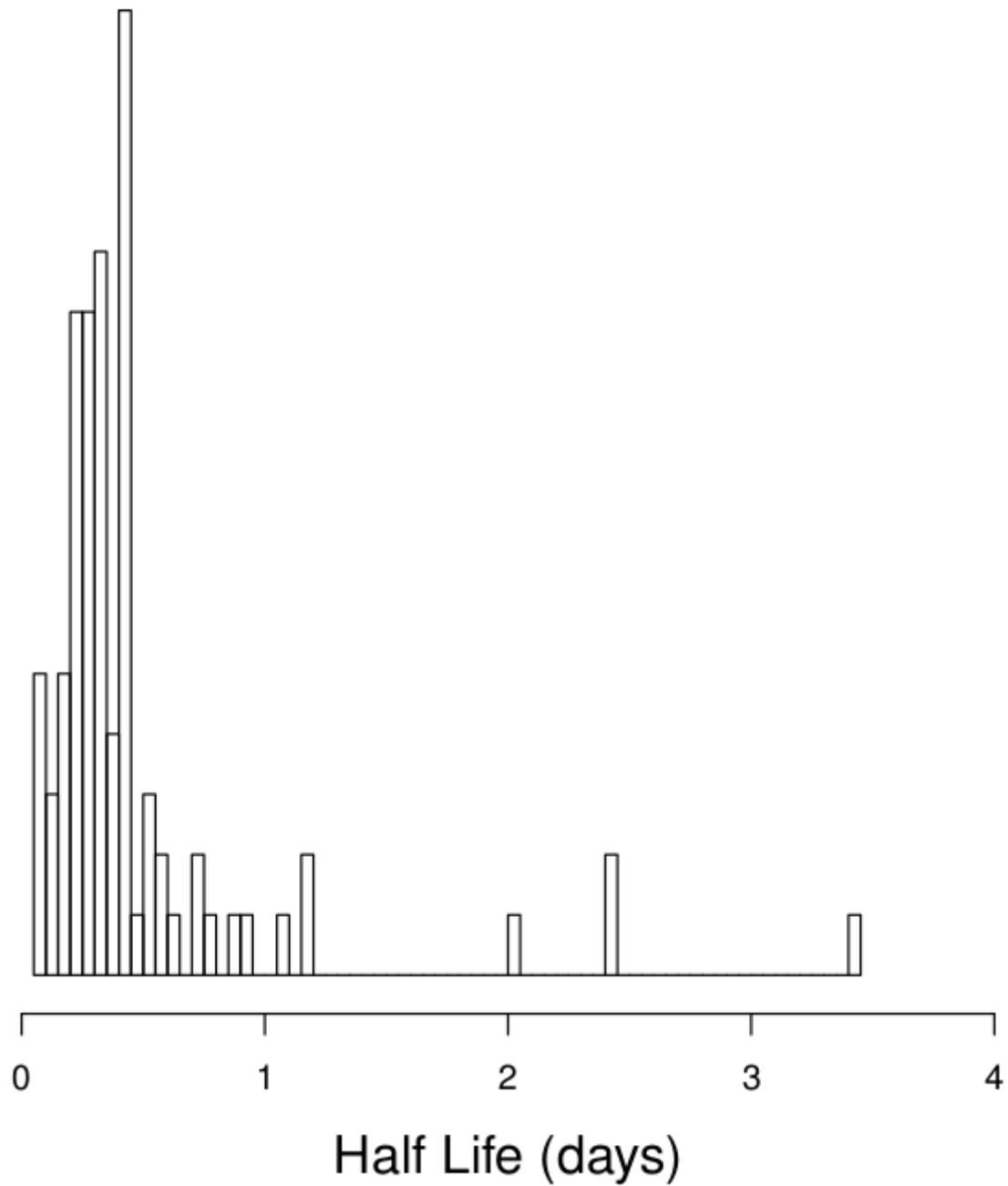

**Figure S3:** Histogram of the half life of each of the 86 chemicals, calculated from the total body clearance rate constant predicted by ACD/Labs.

Figure S4: Internal concentration *vs.* exposure rate for of each of the 86 EDC studied.

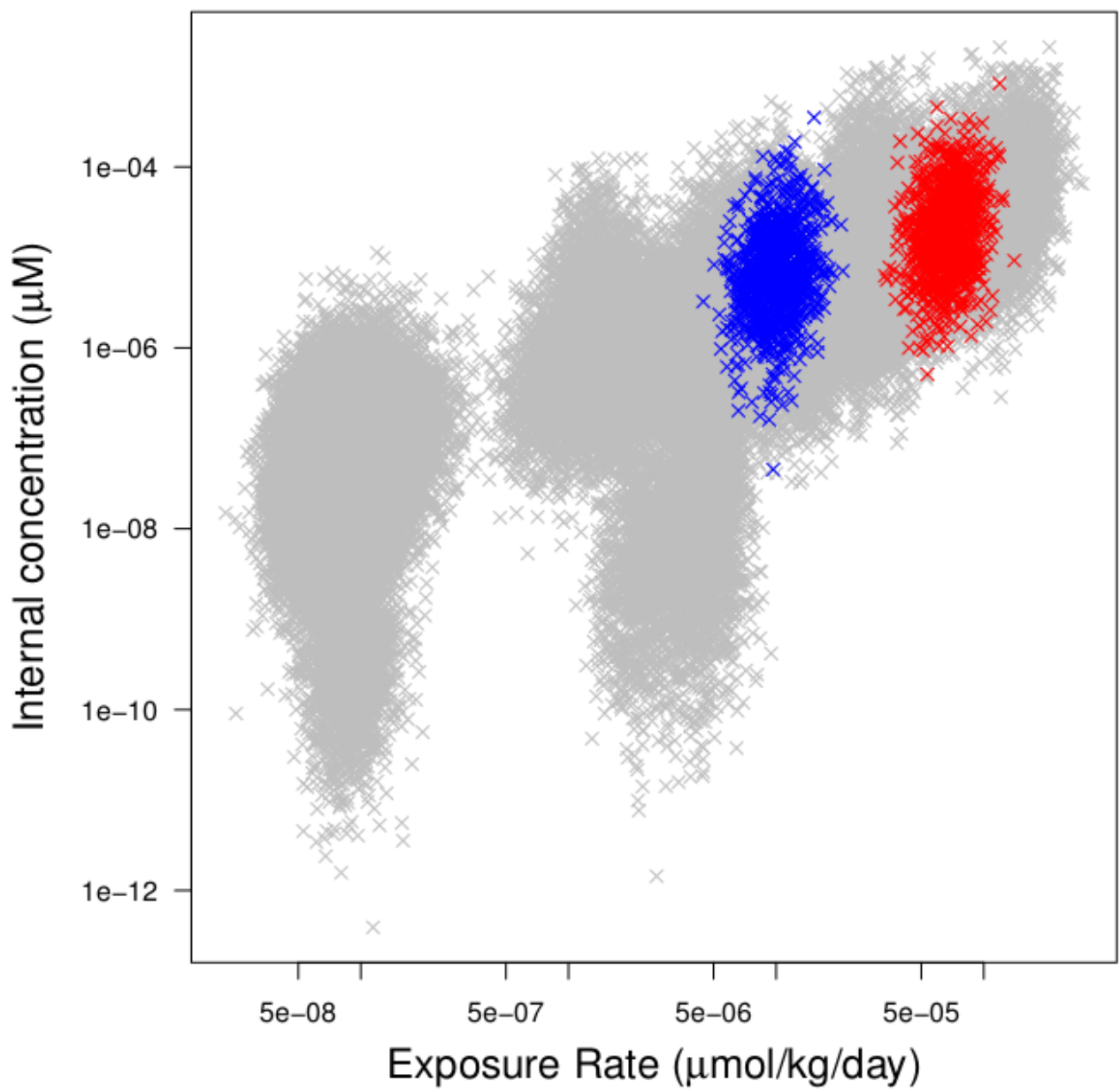

**Figure S4:** Steady-state internal concentrations of the 86 chemicals studied as a function of exposure levels. Exposures and pharmacokinetic parameters were randomly sampled 10000 times for each chemical (see text), resulting in visible spread. Results for two of the chemicals are highlighted (anastrozole in blue and 4,4'-oxydianiline in red).

Figure S5: Example of a random internal exposure profile.

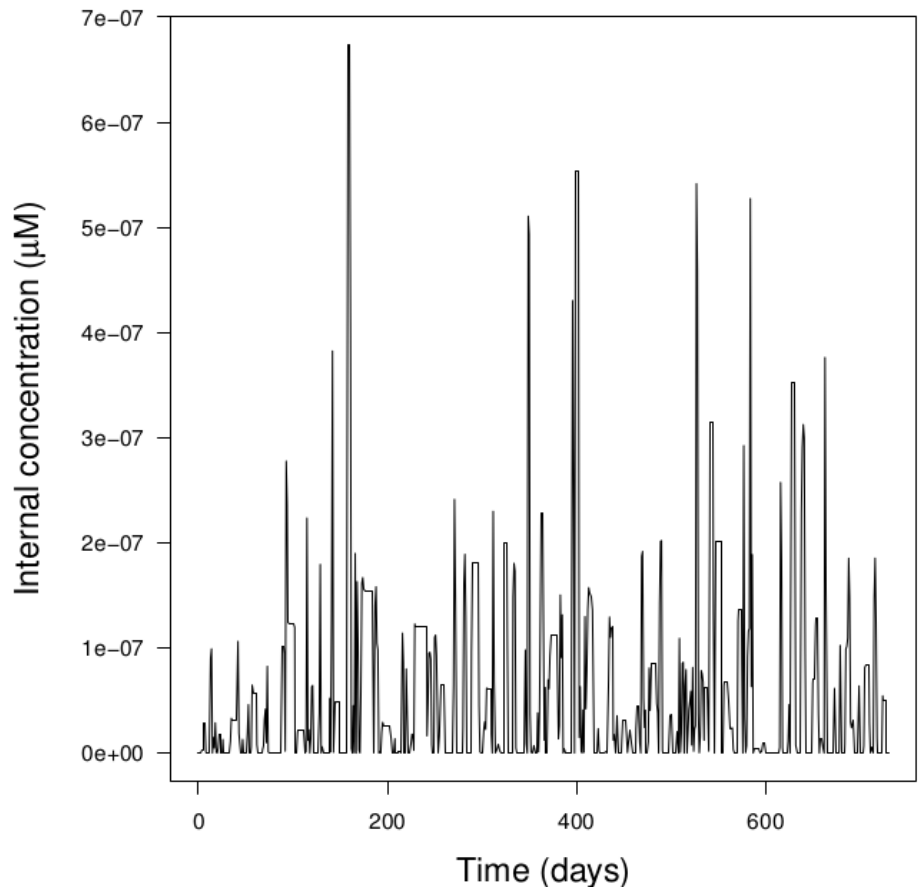

**Figure S5:** Example of a random internal exposure profile (for lindane). Exposure times and rates were Monte Carlo sampled. Internal exposure as a function of time was obtained by numerical integration of the one-compartment PK model.

**Methods S2:** Menstrual cycle model equations.

The equations of the menstrual cycle are adapted from Chen and Ward (2013), as indicated.

The differential equation for the dominant follicle mass (*F*) is a function of a background growth rate, a growth rate component stimulated by *FSH* according to Hill's model, and a degradation rate under *LH* control, also according to Hill's model:

$$\frac{\partial F}{\partial t} = \gamma_F + \frac{\beta_F \cdot F \cdot FSH^{m_6}}{EC_{FSH,F}^{m_6} + FSH^{m_6}} - \frac{\gamma_{F,R} \cdot F^2 \cdot LH^{m_7}}{EC_{LH,F}^{m_7} + LH^{m_7}} \tag{1}$$

The ruptured follicle or scar mass (*R*) is formed from *F* degradation, with first and converted to *corpus luteum* (*C*):

$$\frac{\partial R}{\partial t} = \frac{\gamma_{F,R} \cdot F^2 \cdot LH^{m_7}}{EC_{LH,F}^{m_7} + LH^{m_7}} - \gamma_{R,C} \cdot R \tag{2}$$

The differential equation for *corpus luteum* mass is a function of its formation rate from *R*, and first order degradation:

$$\frac{\partial C}{\partial t} = \gamma_{R,C} \cdot R - \delta_C \cdot C \tag{3}$$

The initial quantities of estradiol (*E2*, in nanograms) and progesterone (*P4*, in nanomoles) in blood and ovaries are given by:

$$Q_{E2,b}(t=0) = 60 \cdot V_{blood} \tag{4}$$

$$Q_{P4,b}(t=0) = 10 \cdot V_{blood} \tag{5}$$

$$Q_{E2,o}(t=0) = 4000 \cdot V_{ovary} \tag{6}$$

$$Q_{P4,o}(t=0) = 1200 \cdot V_{ovary} \tag{7}$$

The concentration of *E2* (nanograms/L) and *P4* (nanomoles/L) in blood and ovaries are:

$$C_{E2,b} = Q_{E2,b} / V_{blood} \tag{8}$$

$$C_{E2,o} = Q_{E2,o} / V_{ovary} \tag{9}$$

$$C_{P4,b} = Q_{P4,b} / V_{blood} \tag{10}$$

$$C_{P4,o} = Q_{P4,o} / V_{ovary} \tag{11}$$

*GnRH* (arbitrary units) is promoted by blood *E2*, and inhibited by blood *P4*, according to Hill's models:

$$Gn = \frac{E2_b^{m_1}}{EC_{E2,Gn}^{m_1} + E2_b^{m_1}} \cdot \frac{EC_{P4,Gn}^{m_2}}{EC_{P4,Gn}^{m_2} + P4_b^{m_2}} \quad (12)$$

The blood *FSH* concentration (in µg/L) in under positive control by past values of *GnRH* ($\tau_{FSH}$ days before) and negative control by current *E2*, according to Hill's models:

$$FSH = \left(FSH_{min} + \frac{(FSH_{max} - FSH_{min}) \cdot Gn_{(t-\tau_{FSH})}^{m_3}}{EC_{Gn_{FSH}}^{m_3} + Gn_{(t-\tau_{FSH})}^{m_3}}\right) \cdot \frac{EC_{E2_{FSH}}^{m_4}}{EC_{E2_{FSH}}^{m_4} + E2_b^{m_4}} \quad (13)$$

The blood *LH* concentration (in µg/L) in under positive control by past values of *GnRH* ($\tau_{LH}$ days before), according to Hill's model:

$$LH = LH_{min} + \frac{(LH_{max} - LH_{min}) \cdot Gn_{(t-\tau_{LH})}^{m_5}}{EC_{Gn_{LH}}^{m_5} + Gn_{(t-\tau_{LH})}^{m_5}} \quad (14)$$

The differential equation for *E2* quantity in blood is a function of transport to and from the ovaries, extra-ovarian production, and elimination rates:

$$\frac{\partial Q_{E2,b}}{\partial t} = F_{ovary}(C_{E2,o} - C_{E2,b}) + Kp_{E2} \cdot \beta_{E2} - Ke_{E2} \cdot C_{E2,b} \quad (15)$$

The differential equation for *E2* quantity in the ovaries is as a function of its transport to and from blood, and its follicular and luteal production, which can be both affected by endocrine disruption (*ED*):

$$\frac{\partial Q_{E2,o}}{\partial t} = ED_{cyp19} \cdot Kp_{E2} \cdot (\beta_{E2F} \cdot F + \beta_{E2C} \cdot C) - F_{ovary}(C_{E2,o} - C_{E2,b}) \quad (16)$$

The differential equation for the quantity of *P4* in blood is a function of its transport to and from the ovaries, extra-ovarian production, and elimination rates:

$$\frac{\partial Q_{P4,b}}{\partial t} = F_{ovary}(C_{P4,o} - C_{P4,b}) + Kp_{P4} \cdot \beta_{P4} - Ke_{P4} \cdot C_{P4,b} \quad (17)$$

The differential equation for the quantity of *P4* in the ovaries is a function of transport to and from blood, and luteal production:

$$\frac{\partial Q_{P4,o}}{\partial t} = Kp_{P4} \cdot \beta_{P4C} \cdot C - F_{ovary}(C_{P4,o} - C_{P4,b}) \quad (19)$$

**Table S3:** Menstrual cycle parameter values. Value are from Chen and Ward (2013), except were indicated.

| Parameter | Symbol | Value (units) |
|---|---|---|
| Ratio of disrupted over control *E2* synthesis | $ED_{CYP19}$ | - [a] |
| Time delay for *GnRH* control of *FSH* | $\tau_{FSH}$ | 1 (day) |
| Time delay for *GnRH* control of L*H* | $\tau_{LH}$ | 1.4 (day) |
| Baseline *E2* concentration | $\beta_{E2}$ | 0.1 (ng/L) |
| Follicular contribution to *E2* synthesis | $\beta_{E2,F}$ | 0.48 (ng/L/mass of *F*) |
| Luteal contribution to *E2* synthesis | $\beta_{E2,C}$ | 0.36 (ng/L/mass of *F*) |
| *E2* clearance from blood | $K_{e,E2}$ | 1600 (L/day) |
| *E2* synthesis scale factor | $K_{p,E2}$ | 1600 (L/day) |
| Baseline *P4* concentration | $\beta_{P4}$ | 0 (nmol/L) |
| Luteal contribution to *P4* synthesis | $B_{P4,C}$ | 0.12 (nmol/L/mass of *C*) |
| *P4* clearance from blood | $K_{e,P4}$ | 3621 (L/day) [b] |
| *P4* synthesis scale factor | $K_{p,P4}$ | 3621 (L/day) |
| *E2* $EC_{50}$ for *GnRH* secretion promotion | $EC_{E2,Gn}$ | 185 (ng/L) |
| *E2* Hill exponent for *GnRH* secretion | $m_1$ | 2 (unitless) |
| *P4* $EC_{50}$ for *GnRH* secretion inhibition | $EC_{P4,Gn}$ | 10 (nmol/L) |
| *P4* Hill exponent for *GnRH* secretion inhibition | $m_2$ | 2 (unitless) |
| *GnRH* $EC_{50}$ for *FSH* secretion promotion | $EC_{Gn,FSH}$ | 0.453 (arbitrary units) |
| *GnRH* Hill exponent for *FSH* secretion promotion | $m_3$ | 2 (unitless) |
| *E2* $EC_{50}$ for *FSH* secretion inhibition | $EC_{E2,FSH}$ | 432 (ng/L) |
| *E2* Hill exponent for *FSH* secretion inhibition | $m_4$ | 4 (unitless) |
| Minimum *FSH* concentration | $FSH_{min}$ | 28.5 (μg/L) |
| Maximum *FSH* concentration | $FSH_{max}$ | 500 (μg/L) |
| *GnRH* $EC_{50}$ for *LH* release | $EC_{Gn,LH}$ | 0.447 (arbitrary units) |
| *GnRH* Hill exponent for *LH* release | $m_5$ | 10 (unitless) |
| Minimum *LH* concentration | $LH_{min}$ | 36.4 (μg/L) |
| Maximum *LH* concentration | $LH_{max}$ | 676 (µg/L) |
| Baseline follicle growth rate | $\gamma_F$ | 0 (arbitrary units/day) |
| *FSH* $EC_{50}$ for follicle growth stimulation | $EC_{FSH,F}$ | 150 (µg/L) |
| *FSH* Hill exponent for follicle growth stimulation | $m_6$ | 2 (unitless) |
| *FSH* enhanced follicle growth stimulation rate constant | $\beta_F$ | 2.1 (1/day) |
| *LH* $EC_{50}$ for follicle rupture | $EC_{LH,F}$ | 325 (µg/L) |
| *LH* Hill exponent for follicle rupture | $m_7$ | 2 (unitless) |
| Follicle rupture rate constant per unit follicle mass | $\gamma_{F,R}$ | 0.06 (1/day/arbitrary units) |
| *R* to *C* conversion rate constant | $\gamma_{R,C}$ | 0.285 (1/day) |
| *Corpus luteum* decay rate constant | $\delta_C$ | 0.345 (1/day) |
| Blood volume | $V_{blood}$ | 4.1 (L) [c] |
| Ovaries volume | $V_{ovary}$ | 0.2 (L) [c] |
| Ovaries blood flow | $F_{ovary}$ | 30 (L/day) [c] |

[a] The value of this parameter depends on exposure to endocrine disrupting chemicals.

[b] Little et al. (1966).

[c] International Commission on Radiological Protection (ICRP) (2002).

**Table S4:** Menstrual cycle initial state variables' values.

| State variable | Symbol | Value (units) |
| --- | --- | --- |
| Dominant follicle mass | $F$ | 79.48 (arbitrary units) |
| Ruptured follicle (scar) mass | $R$ | 20.23 (arbitrary units) |
| *Corpus luteum* mass | $C$ | 33.35 (arbitrary units) |
| Quantity of estradiol in blood | $Q_{E2,b}$ | 206.1 (ng) |
| Quantity of estradiol in ovaries | $Q_{E2,o}$ | 545.1 (ng) |
| Quantity of progesterone in blood | $Q_{P4,b}$ | 16.43 (nmol) |
| Quantity of progesterone in ovaries | $Q_{P4,o}$ | 97.53 (nmol) |

**Figure S6**: Simulation of normal ovarian cycling and comparison to data.

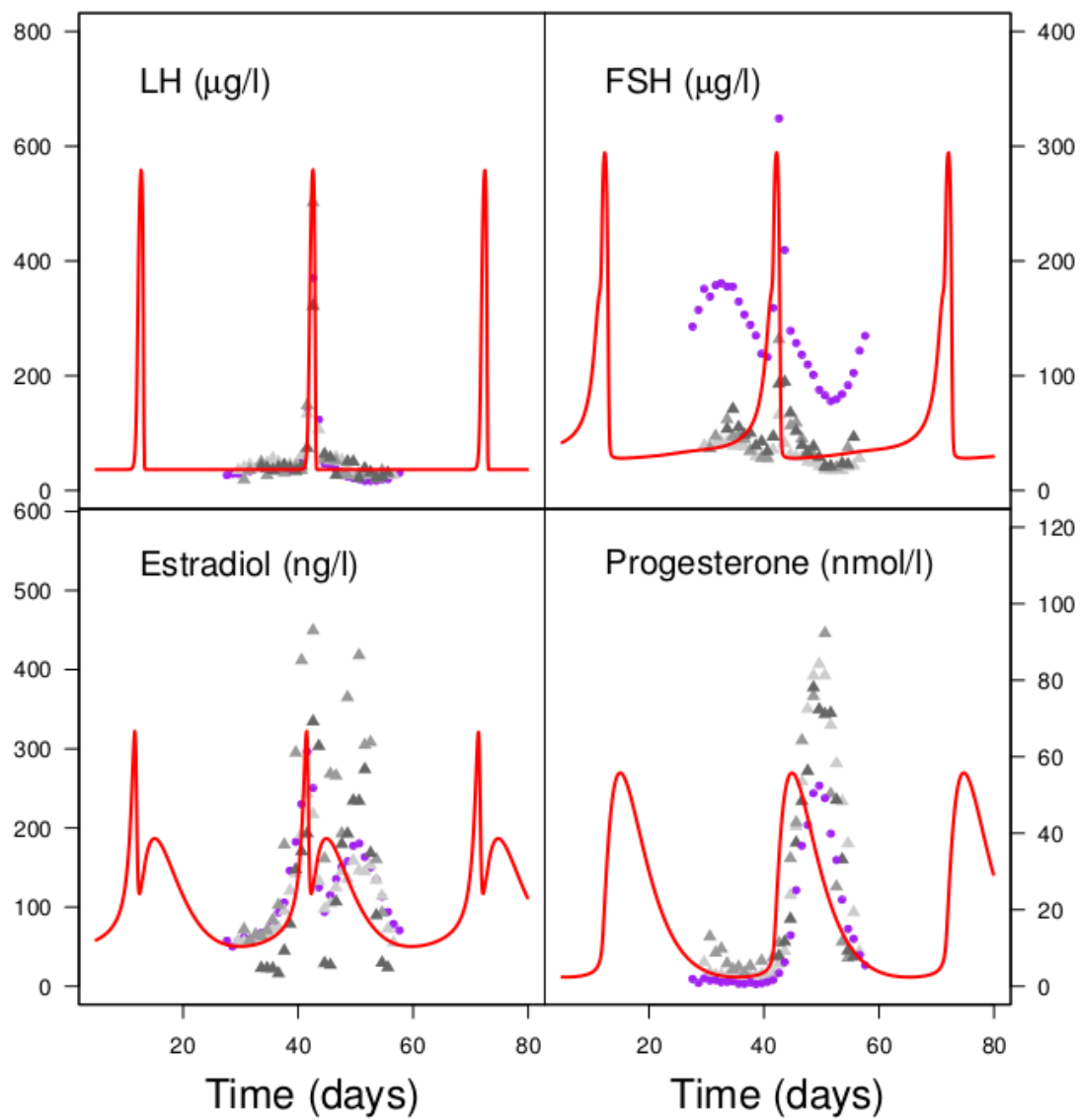

**Figure S6:** Model simulated time course of *LH*, *FSH*, *E2* and *P4* during normal human ovarian cycles (red lines). The model equations and parameter values used are given in Methods S2 and Tables S3 and S4 above. Data are overlaid as points: purple circles from McLachlan et al. (1990, Fig. 1); gray triangles from Welt et al. (1999). The LH data points from Welt *et al.* were rescaled so that their baseline values match with those of McLachlan et al. The same rescaling factor was applied to Welt et al. FSH data.

**Figures S7-S10**: Simulation of normal and perturbed ovarian cycling.

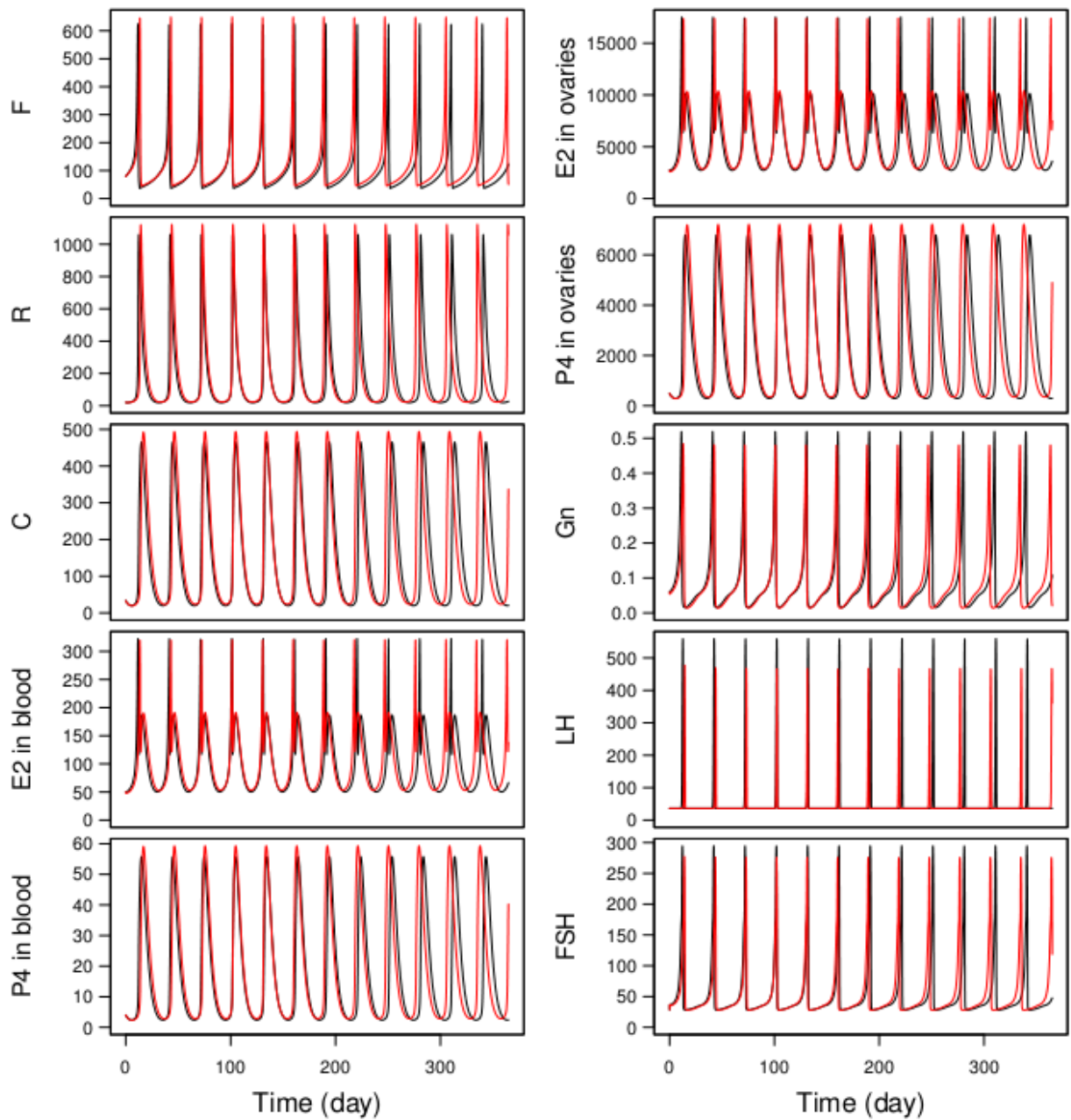

**Figure S7:** Simulated time course of all model variables during normal ovarian cycles (parameter $ED_{CYP19}$ set to 1, black lines) and with a 5% constant inhibition of aromatase (parameter $ED_{CYP19}$ set to 0.95). The model equations and parameter values used are given in Methods S2 and Tables S3 and S4 above. *LH* and *FSH* units are in µg/L. *GnRH* units are arbitrary, the units of the other variables are given in Table S4 above. The perturbed cycle differs only slightly from normal.

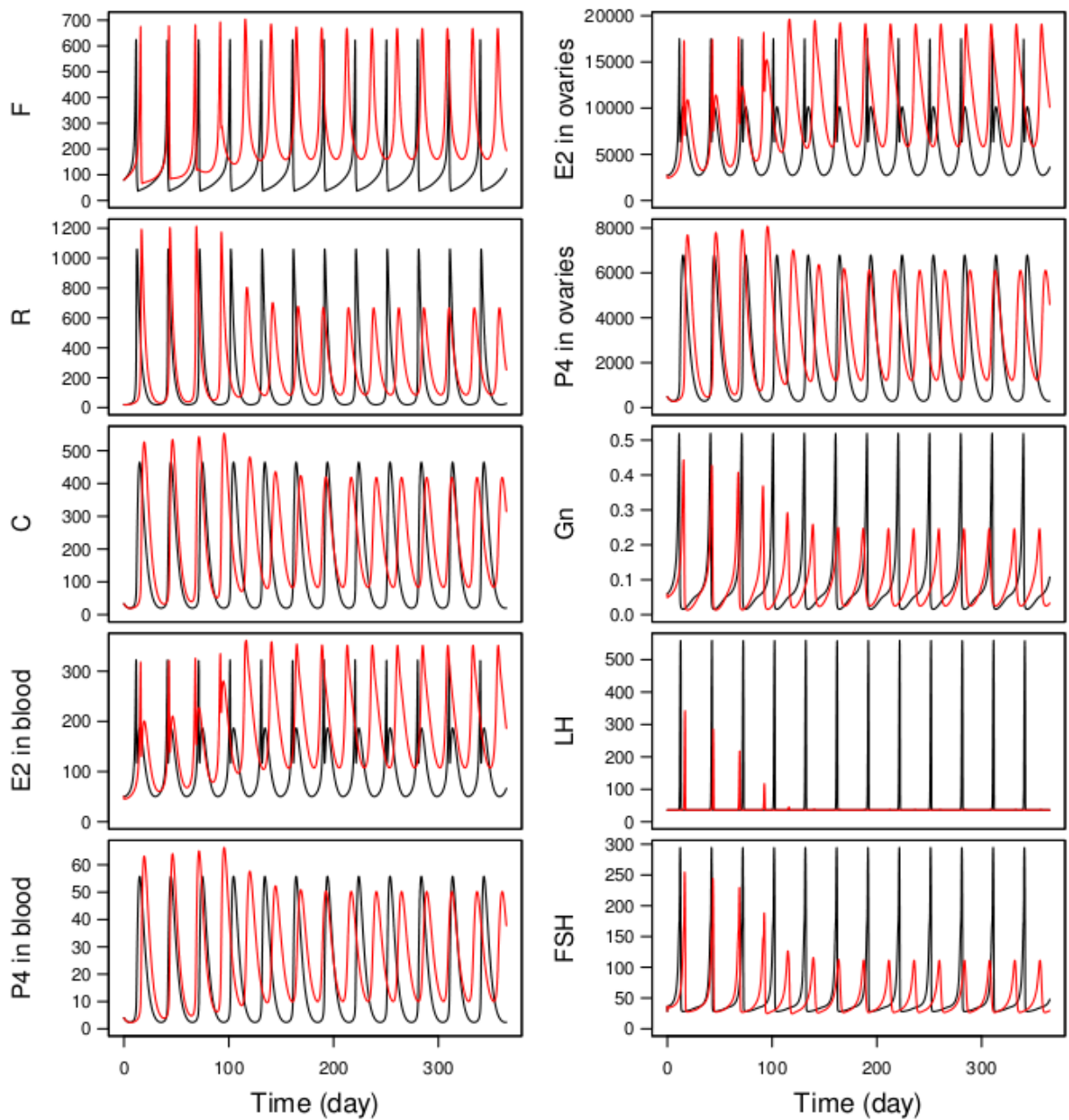

**Figure S8:** Simulated time course of all model variables during normal ovarian cycles (parameter $ED_{CYP19}$ set to 1, black lines) and with A 10% constant inhibition of aromatase (parameter $ED_{CYP19}$ set to 0.9). The model equations and parameter values used are given in Methods S2 and Tables S3 and S4 above. *LH* and *FSH* units are in µg/L. *GnRH* units are arbitrary, the units of the other variables are given in Table S4 above. LH peaks disappear rapidly and the systems bifurcates to a new solution.

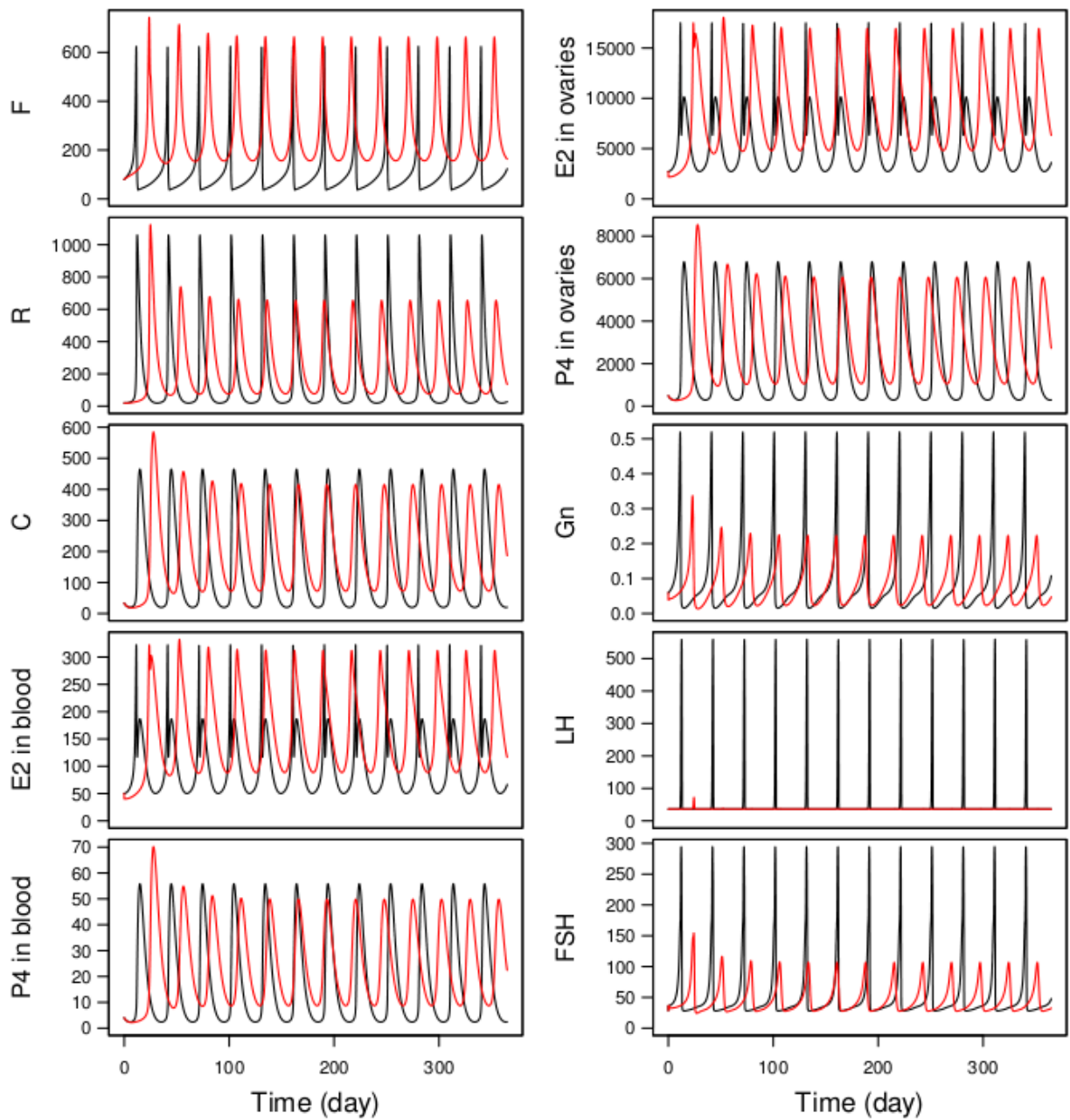

**Figure S9:** Simulated time course of all model variables during normal ovarian cycles (parameter $ED_{CYP19}$ set to 1, black lines) and with A 20% constant inhibition of aromatase (parameter $ED_{CYP19}$ set to 0.8). The model equations and parameter values used are given in Methods S2 and Tables S3 and S4 above. *LH* and *FSH* units are in μg/L. *GnRH* units are arbitrary, the units of the other variables are given in Table S4 above. The cycle is profoundly disrupted.

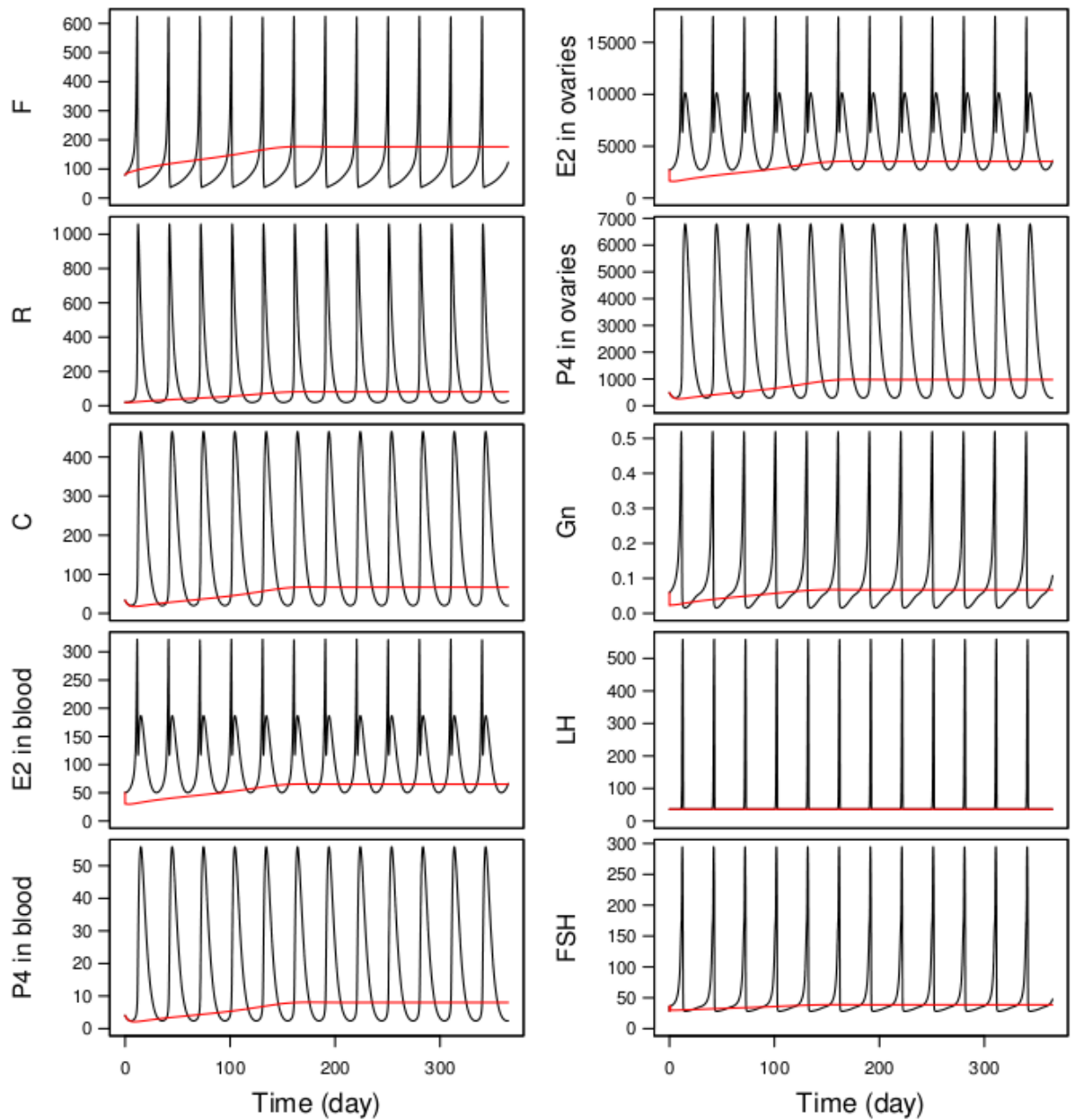

**Figure S10:** Simulated time course of all model variables during normal ovarian cycles (parameter *ED$_{CYP19}$* set to 1, black lines) and with A 40% constant inhibition of aromatase (parameter *ED$_{CYP19}$* set to 0.6). The model equations and parameter values used are given in Methods S2 and Tables S3 and S4 above. *LH* and *FSH* units are in µg/L. *GnRH* units are arbitrary, the units of the other variables are given in Table S4 above. The cycle is totally disrupted.

**Figure S11**: Simulation of estradiol time profiles during time-varying exposures to 86 EDCs.

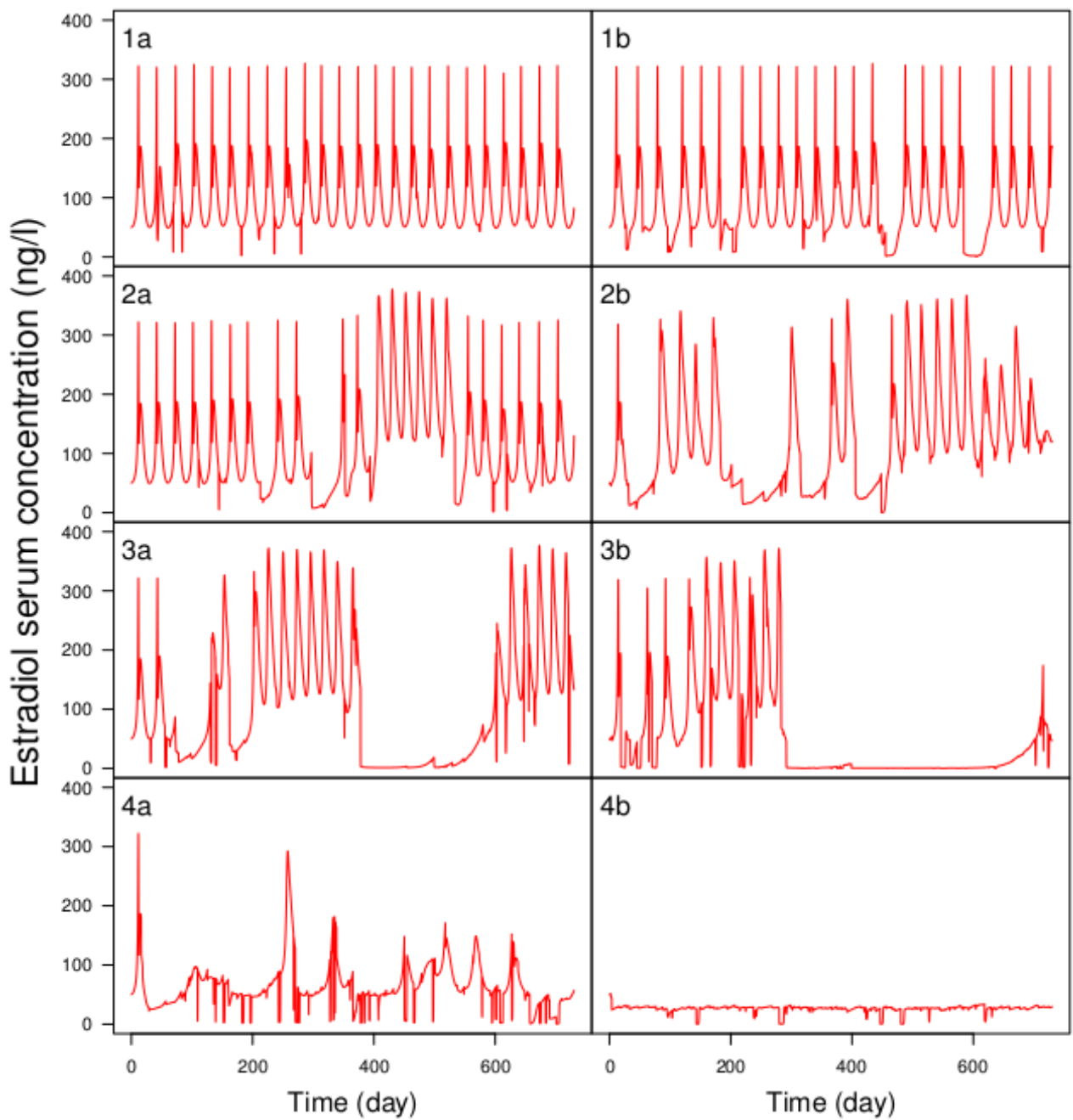

**Figure S11:** Typical simulated time profiles of estradiol concentration during time-varying exposures to random mixtures of 86 aromatase inhibitors. Four classes (rows) of increasing disruption are illustrated (see main text). Left column: least disrupted profile in its class; Right column: most disrupted. Panel 1a shows a practically normal profile with regular ovulation peaks. Panel 4b shows complete disruption.